\documentclass[USenglish,oneside,twocolumn]{article}

\usepackage[utf8]{inputenc}
\usepackage[big]{dgruyter_NEW}

\cclogo{\includegraphics{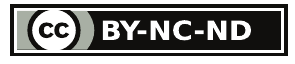}}

\usepackage{amsmath,amssymb}
\usepackage[ruled,vlined,linesnumbered]{algorithm2e} 
\usepackage{graphicx}
\usepackage{xspace}
\usepackage{color}
\usepackage{stmaryrd}
\usepackage{mathtools}
\usepackage{url}
\usepackage{caption}
\usepackage{subcaption}
\usepackage{tikz}
\usepackage{enumitem}
\usepackage{array}
\usepackage{multirow}

\newcommand*\rot{\rotatebox{90}}

	\newcommand{\red}[1]{{\color{red}{#1}}}

	\newcommand*\circled[1]{\protect\tikz[baseline=(char.base)]{
    		        \protect\node[shape=circle,draw,inner sep=2pt] (char) {#1};}}

\newtheorem{theorem}{Example}

\usepackage[font=small,justification=centering,labelfont=bf,tableposition=top]{caption}
\DeclareCaptionLabelFormat{andtable}{Figure#1~#2  \&  Table~\thetable}
\cclogo{\includegraphics{by-nc-nd.pdf}}
\begin{document}

\title{Near-Optimal Fingerprinting with Constraints}

\author*[1]{G{\'a}bor Gy{\"o}rgy Guly{\'a}s}

\author[2]{Gergely Acs}

\author[3]{Claude Castelluccia}

\affil[1]{INRIA, E-mail: gabor.gulyas@inria.fr}

\affil[2]{INRIA, E-mail: gergely.acs@inria.fr}

\affil[3]{INRIA, E-mail: claude.castelluccia@inria.fr}

\begin{abstract}
{Several recent studies have demonstrated that people show large behavioural uniqueness. This has serious privacy implications as most individuals become increasingly re-identifiable in large datasets or can be tracked, while they are browsing the web, using only a couple of their attributes, called as their fingerprints. Often, the success of these  attacks depends on explicit constraints on the number of attributes learnable about individuals, i.e., the size of their fingerprints. These constraints can be budget as well as technical constraints imposed by the data holder. For instance, Apple restricts the number of applications that can be called by another application on iOS in order to mitigate the potential privacy threats of leaking the list of installed applications on a device. 
In this work, we address the problem of identifying the attributes (e.g., smartphone applications) that can serve as a fingerprint of users given constraints on the size of the fingerprint. We give the best fingerprinting algorithms in general, and evaluate their effectiveness on several real-world datasets. Our results show that current privacy guards limiting the number of attributes that can be queried about individuals is insufficient to mitigate their potential privacy risks in many practical cases.}
\end{abstract}

  \journalname{Proceedings on Privacy Enhancing Technologies}
  \DOI{10.1515/popets-2016-0051}
  \startpage{1}
  \received{2016-02-29}
  \revised{2016-06-02}
  \accepted{2016-06-02}

  \journalyear{}
  \journalvolume{2016}
  \journalissue{4}

\maketitle

\section{Introduction}
\label{sec:intro}
People exhibit large behavioural uniqueness. The way how they move around~\cite{Nature13}, purchase goods~\cite{Science15}, configure their browser~\cite{pan10}, or browse the web~\cite{Olejnik13} make them unique in very large populations.
Even various combinations of seemingly innocuous behavioural attributes, such as in which shops they used to buy goods or where they live and work make them unique \cite{Zang2011}.
For example, it has been demonstrated that only four spatio-temporal positions are enough to uniquely identify an individual 95\% of the times in a dataset of one and a half million users \cite{Nature13}. 

As many devices record people's behavioural patterns today, it becomes relatively easy for third parties to access such personal information. This has detrimental effect on individuals' privacy, as behavioural attributes can be easily used for the purpose of re-identification and tracking in different contexts.
For example, advertiser companies often use behavioural attribute values (e.g., the list of installed fonts \cite{pan10} or applications \cite{twitter14}) to fingerprint website visitors in order to show them more personalized ads. That is, each time a user visits a website where a 3rd party tracker is embedded, the tracker retrieves the user's fingerprint (such as certain fonts installed on the device \cite{cross12}) and \emph{link} all his visited sites together \cite{Acar:2014}. As visited sites can belong to very different topics, the tracker can build accurate interest profiles of visitors in order to show them more personalized ads. However, the list of visited sites can also potentially uncover sensitive information such as sexual behaviour or religion among others. 
Indeed, most popular
website categories that employ such user tracking were found to be
porn (15\%) and dating (12.5\%) \cite{Bujlow:2015}.

Several countermeasures have been introduced against fingerprinting in the context of web-tracking \cite{Bujlow:2015}. A simple solution is to completely disable all techniques which allow a tracker to fingerprint visitors, such as the execution of Javascript and Flash programs in a browser. However, this causes significant degradation of user experience and hence it is a less appealing option. Another, more flexible approach is to enable but \emph{constrain} the usage of such techniques. For example, only \emph{limited number} of system fonts can be used/queried by an entire website \cite{tord15}, or only \emph{limited number} of applications can be opened from another application on a smartphone \cite{appleios9}.
Hence, the goal of an adversary is to identify at most a given number of attributes (e.g., fonts or applications), whose values make either a \emph{targeted} user or \emph{every} user as unique as possible in the dataset, i.e., the number of users sharing the same values of these attributes is minimized.

In this paper, we show that the practice of limiting the number of  attribute queries per user to mitigate the privacy risks of fingerprinting is often inadequate. We construct efficient greedy fingerprinting algorithms given constraints on the maximum size of the fingerprint. We provide essentially the best fingerprinting algorithms under standard computational assumptions. In particular, we address two general problems. Suppose a positive integer $s$. 
First, given a targeted user and all his attribute values, we want to know the subset of attributes with cardinality at most $s$, called the individual fingerprint of the user, whose values make this user as unique as possible in a given dataset. This is referred to as the \emph{targeted fingerprinting problem} and analogous to the re-identification (de-anonymization) problem when, for example, a user of the Tor Browser is re-identified using some of his/her attributes which can be queried by visited websites. Second, we want to know the subset of attributes with cardinality at most $s$, called general fingerprint, whose values make each member of a group as unique as possible within the group. This is referred to as the  \emph{general fingerprinting problem} and analogous to the linking problem when, for example, the browsing activities of users are tracked, perhaps without re-identifying them.

For example, in Table \ref{tab:ex}, the individual fingerprint of $U_1$ with maximum size 2 is $\{ A_2, A_4\}$, and the fingerprint value is $\{A_2 :  0, A_4: 1\}$ which make $U_1$ unique among 5 other users. However, $U_2$ has many possible fingerprints with maximum size 2 (e.g.,  $\{A_2, A_4\}$), but none of them provides a unique fingerprint value to $U_2$, because there is always at least another user which remains indistinguishable from $U_2$ with respect to the fingerprint.
Table \ref{tab:groups} shows the anonymity sets of 
all possible subsets of attributes with cardinality 3 (in one anonymity set, users share the same values of these attributes). We find that the only general fingerprint of all the six users, with maximum size 3,  is $\{A_2, A_3, A_4\}$, because this is the combination of attributes (with size at most 3) which yields the smallest anonymity sets. 

\begin{table}
\small
\centering
\begin{tabular}{|c|c|c|c|c|}
\hline
\emph{User \#} & $\mathbf{A_1}$ & $\mathbf{A_2}$ & $\mathbf{A_3}$ & $\mathbf{A_4}$ \\
\hline \hline
\emph{$U_1$} &  1 & 0 & 1 & 1\\ \hline
\emph{$U_2$} &  1 & 1 & 1 & 1\\ \hline
\emph{$U_3$} &  0 & 1 & 0 & 1\\ \hline
\emph{$U_4$} &  1 & 0 & 1 & 0\\ \hline
\emph{$U_5$} &  1 & 1 & 1 & 0\\ \hline
\emph{$U_6$} &  1 & 1 & 0 & 0\\ \hline
\end{tabular}
\caption{Example dataset where each row is a record of a user with binary attributes $\mathbf{A_1}, \mathbf{A_2}, \ldots, \mathbf{A_4}$. \label{tab:ex}}
\end{table}

\begin{table*}
\small
\centering
\begin{tabular}{|c|l|}
\hline
\emph{Subset of Attributes} & \emph{Anonymity sets} \\ \hline
\hline
$\{ \mathbf{A_1}, \mathbf{A_2}, \mathbf{A_3} \}$ &  $\{U_1, U_4\}$, $\{U_2, U_5\}$,  $\{U_3\}$, $\{U_6\}$\\ 
\hline
$\{ \mathbf{A_1}, \mathbf{A_2}, \mathbf{A_4} \}$ & $\{U_1\}$, $\{U_2\}$, $\{U_3\}$, $\{U_4\}$, $\{U_5, U_6\}$ \\
\hline 
$\{\mathbf{A_1}, \mathbf{A_3}, \mathbf{A_4} \}$&  $\{U_1, U_2\}$, $\{U_3\}$, $\{U_4, U_5\}$, $\{U_6\}$\\ 
\hline
\red{\small $\{\mathbf{A_2}, \mathbf{A_3}, \mathbf{A_4} \}$} &  \red{ \small $\{U_1\}$, $\{U_2\}$, $\{U_3\}$, $\{U_4\}$, $\{U_5\}$, $\{U_6\}$ } \\ \hline
\end{tabular}
\caption{The anonymity sets of all possible subsets of attributes (from Table \ref{tab:ex}) with size 3. The single general fingerprint is $\{\mathbf{A_2}, \mathbf{A_3}, \mathbf{A_4}\}$ whose values make every user unique.}
\label{tab:groups}
\end{table*}

We believe that the solutions of the above fingerprinting problems also have great importance in practice due to the upcoming European General Data Protection Regulation which mandates privacy risk assessment \cite{GDPR}. This requires to measure the re-identification risk given certain ``reasonable'' constraints on the adversarial background knowledge. Moreover, the problem of user fingerprinting indirectly also appears in the Article 29 Working Party Opinion 05/2014 on data anonymization techniques \cite{Art29op} which describes the linkability of records concerning the same data subject in an anonymized dataset as a privacy weakness.

Finally, we must note that our problems are different from the one in \cite{masking_vldb}, which is about finding a minimal subset of attributes that provides a certain separation of all the users in a dataset. That is, given a separation value $\alpha$, the minimum subset of attributes which provides separation $\alpha$ (i.e., it separates at least $\alpha$ fraction of all possible record pairs) is approximated in \cite{masking_vldb}. By contrast, in our problems,  we are given a threshold $s$ and we aim to give a subset of attributes with size at most $s$ which provides the best separation. Obviously, the larger $s$ the better separation is provided. 
Moreover, as we describe in Section \ref{sec:gen_fp}, our general fingerprinting algorithm solves some problems in \cite{masking_vldb} more efficiently than prior algorithms.
We also note that the naive approach of using a limited number of \emph{random} attributes as a fingerprint, which is also employed by prior re-identification studies \cite{appuni15} \cite{Nature13} \cite{Science15}, is inferior to our solution. In particular, these naive approaches build individual fingerprints from only those attributes whose values are set in a record. For example, in Table \ref{tab:ex}, $\{ A_2, A_4\}$ is not a possible fingerprint of $U_1$ with such a naive approach, as $A_2$ is not set in the first record. In fact, $U_1$ has no unique fingerprint with size 2 out of the attributes whose values are 1 in $U_1$'s record. Even more, in our work, missing attributes such as $\{ A_1 = 0\}$ could be a valid individual fingerprint of $U_3$.

Our main contributions are summarized as follows:
\begin{itemize}
\item We provide essentially the best fingerprinting algorithms when constraints of the size of the fingerprint are provided. In particular, we first prove that both targeted and general fingerprinting are $\mathbf{NP}$-hard in general and the best possible polynomial time approximations are greedy heuristics.
\item We provide real-life applications where these fingerprinting methods can be used in practice. In particular, we analyze the privacy protection used by many popular services, such as iOS or Tor Browser Bundle, which are based on limiting the number items (e.g., installed apps or fonts) that can be queried on a device for the purpose of fingerprinting. Our aim is to raise awareness about the weakness of such privacy guards. 
\item We evaluate our fingerprinting algorithms on different large real-life datasets. Specifically, we analyze the separation of different individual as well as general fingerprints on several datasets, where a record can contain the list of installed fonts, smartphone applications, or visited locations by an individual.
\end{itemize} 
The code base related to this paper is available at \url{https://github.com/gaborgulyas/constrainted_fingerprinting}.

\section{Model}
\label{sec:model}
Suppose that each individual in a population $P$ has a set of items from a larger universe $\mathcal{U}$ with size $n$. 
The universe can contain any observable items of individuals (e.g., set of all possible applications installed on individuals' smartphones, set of possible visited locations, etc.).
Therefore, each user is represented by a binary vector $u$, called the user's profile, where $u[i] = 1$ only if the user has the corresponding item $i \in \mathcal{U}$. 

Our goal is twofold. First, we want to identify \emph{at most} $s$ items, called individual fingerprint, which separate a single individual from others the most, i.e., whose values provide the largest uniqueness to the individual in the population. We refer to this problem as \emph{targeted fingerprinting}. Second,  we want to identify \emph{at most} $s$ items, called general fingerprint, which simultaneously separate all users from each other in a group the most, i.e., whose values make each member of the group as unique as possible within the group.
We call this problem as \emph{general fingerprinting}. 

In particular, we are given the following constraints. (1) We have only access to a subset $B$ of the whole population $P$, where one can think of $B$ as a subsample of $P$. (2) We can query at most $s$ items of a user to check whether the user has these items in his/her profile or not. Then, our exact problems to be solved are as follows.

\begin{description}
\item[Targeted fingerprinting:]  Given a targeted profile $u$ and a subset $B$ of population $P$. Identify at most $s$ items $\{i_1, i_2, \ldots, i_s\} \subset \mathcal{U}$ such that the number of profiles in $B$ which share identical values at positions $i_1, i_2, \ldots, i_s$ with $u$ is minimized.

\item[General fingerprinting:]  Given a subset $B$ of population $P$. Identify at most $s$ items $\{i_1, i_2, \ldots, i_s\} \subset \mathcal{U}$ such that the number of profiles in $B$ which share identical values at positions $i_1, i_2, \ldots, i_s$ is minimized.
\end{description}

Simply put, we intend to compute the uniqueness of different itemsets in $B$ and generalize these results to the whole population.
The underlying assumption of this approach is that $B$ is a subset of the population $P$, and hence if $B$ has sufficient number of users, the uniqueness of different itemsets in $B$ and $P$ are also similar.

\section{Hardness of fingerprinting}
\label{sec:hardness}
We show that identifying the fingerprint items for both problems described in Section \ref{sec:model} is $\mathbf{NP}$-hard. In both cases, we reduce the problem to the Maximum Coverage Problem which is $\mathbf{NP}$-hard \cite{Chekuri04maximumcoverage}. \\ \medskip

\noindent \textbf{Maximum Coverage Problem:} Given a positive integer  $s$  and a collection of sets  $C = C_1, C_2, \ldots, C_m$. Find a subset  $C' \subseteq C$ of sets such that  $| C' | \leq s$ and the number of covered elements  $\left| \bigcup_{C_j \in C'}{C_j} \right|$  is maximized. \medskip

The reductions are as follows. \\ \medskip

\noindent \textbf{Targeted fingerprinting:} Let $u$ denote the profile of the targeted user.
Assign a set of profiles $P_i \subseteq B$ to each item $i \in \mathcal{U}$, where item $i$ of each profile in $P_i$ has value $u[i]$. Let $\mathbb{P} = \{ P_i | i \in \mathcal{U}\}$, where $|\mathbb{P}| = |\mathcal{U}|$. 
The problem is to find at most $s$ sets in $\mathbb{P}$ such that the size of their intersection is minimized. This is identical to finding at most $s$ sets such that the size of the complement of their intersection is maximized. Following from De Morgan's laws, this is further equivalent to finding at most $s$ sets in $\overline{\mathbb{P}} = \{ \overline{P_i} | i \in \mathcal{U}\}$ such that the size of their \emph{union} is maximized, where $\overline{P_i}$ contains all profiles whose value at $i$ is \emph{not} $u[i]$.
This problem is identical to the Maximum Coverage Problem, and hence Problem 1 is $\mathbf{NP}$-hard. \\ \medskip

\noindent \textbf{General fingerprinting:}
Let $V$ denote the number of all pairs of profiles from $B$, where $|V| = {|B| \choose 2}$.  For each item $i \in \mathcal{U}$, let $T_i \subset A$ denote all pairs of profiles which have \emph{different} values at $i$ (i.e., they can be separated by item $i$). Clearly, $\bigcup_{i\in \mathcal{U}} T_i = V$. Let $\mathbb{T} = \{ T_i | i \in \mathcal{U}\}$, where $|\mathbb{T}| = |\mathcal{U}|$. Hence, Problem 2 is equivalent to finding at most $s$ sets in $\mathbb{T}$ such that the size of their union (i.e., the total number of separated pairs of profiles) is maximized, which is again the Maximum Coverage Problem. 

\section{Algorithms}
\label{sec:algos}
As the underlying problems for both fingerprinting problems are $\mathbf{NP}$-hard, we cannot hope for finding the optimal solutions. However, a simple greedy heuristics, which approximates the solution of the Maximum Coverage problem, provides essentially the best possible approximations for our fingerprinting problems.

The greedy algorithm for the Maximum Coverage Problem iteratively picks the set  which covers the largest number of uncovered elements. It can be shown that this algorithm achieves an approximation ratio of $(1 - 1/e) \approx 0.632$ \cite{Nemhauser78}. That is, if $\mathit{OPT}$ denotes the optimal solution (i.e., the maximum number of covered elements by at most $s$ sets), then the greedy algorithm will cover at least $(1 - 1/e)\cdot \mathit{OPT}$ elements.
Moreover, it has also been proven that, unless $\mathbf{P}=\mathbf{NP}$, there is no $1 -1/e - o(1)$ approximation for the Maximum Coverage Problem \cite{Feige98}, which means  that the greedy approach is essentially the best-possible polynomial time approximation for this problem. 

\subsection{Targeted fingerprinting} 
\label{sec:ind_algo}
The first problem is straightforward to approximate with the greedy approach which is described in Algorithm \ref{alg:deanon}. Instead of maximizing the number of users which disagree on the values of fingerprint queries with the targeted user $u$, we minimize the number of users which \emph{agree} on those values with $u$.

We iteratively squeeze the anonymity set of $u$, denoted by $\mathit{anon\_set}$ in Algorithm \ref{alg:deanon}, which is the set of users sharing identical values at all items of the fingerprint queries $K$ with $u$. Initially, $\mathit{anon\_set}$ is composed of all the users in $B$.
Then, in each iteration, we select the item $i_{\mathit{sep}}$ which separates $u$ from the other profiles in the anonymity set the most, i.e., minimizes the number of profiles which agree at all items of $K$ with $u$. 
For this purpose, we pre-compute
a  table, denoted as $\mathit{users}$ in Line 2, which maps each item $i \in \mathcal{U}$ to the list of profiles which has the value of $u[i]$ at position $i$. 
The computation of this table takes time of $O(|B||\mathcal{U}|)$. Then, we can easily identify the most separating item in each iteration by the intersection of
the anonymity set of $u$ and each list of table $\mathit{users}$. Fast implementations of the intersection of sorted integers are described in \cite{LemireBK14}. If users in the anonymity set are no further distinguishable by any item, we terminate and return $K$  as the final fingerprint (in Line 7). Otherwise, we continue the separation as long as $|K| \leq s$.

This solution has a storage complexity of $O(|B||\mathcal{U}|)$ and computational complexity of $O(s |B||\mathcal{U}|)$.

\begin{algorithm}[!ht]
\caption{Greedy Approximation for Targeted Fingerprinting \label{alg:deanon}}

\small
\DontPrintSemicolon
\KwIn{Fingerprint size $s$, Universe of items $\mathcal{U}$, Dataset $B$, Targeted user's profile $u$}
\KwOut{Fingerprint items $K$} 
$K  := \varnothing$\;
$\mathit{users}[i] := \{u'\, |\, u'\in B \wedge u[i] = u'[i]\}$ for all $i \in \mathcal{U}$\;
$\mathit{anon\_set} := B$\;
\Repeat{$|K| = s$ \textbf{or} $|\mathit{anon\_set}| = 1$}
{
\texttt{// if no further separation is possible}\;
\If{ $ \min_{i \in \mathcal{U} \setminus K}\{ |\mathit{users}[i] \cap \mathit{anon\_set}| \} = |\mathit{anon\_set}|$ }
{ $\mathbf{break}$\; }
$i_{sep} := \arg \min_{i \in \mathcal{U} \setminus K}\{ |\mathit{users}[i] \cap \mathit{anon\_set}| \}$\;
$\mathit{anon\_set} := \mathit{anon\_set} \cap \mathit{users}[i_{\mathit{sep}}]$\;
$K:= K \cup \{i_{\mathit{sep}}\}$\;
}
\KwRet{$K$}

\end{algorithm}

\begin{theorem} Suppose that $u=U_5$ in Table \ref{tab:ex} and $s=3$. First, we select the attribute which makes $U_5$ the most unique. $U_5$ shares the same values of $A_1$, $A_2$, $A_3$ and $A_4$ with 4, 3, 3, and 2 other users, respectively, hence we select $A_4$ in the first iteration. As a result, $U_4$ and $U_6$ still remain indistinguishable from $U_5$. In the next iteration,  we select the attribute which distinguishes $U_5$ from $U_4$ and $U_6$ the most. We have two options: $A_2$ or $A_3$, as $A_1$ are identical for all of them. We select, say, $A_2$, which means that $U_5$ and $U_6$ are still indistinguishable. Hence, in the final iteration, we select $A_2$, which yields $\{A_2, A_3, A_4\}$ as the individual fingerprint of $U_5$. 
\end{theorem}

\subsection{General fingerprinting} 
\label{sec:gen_fp}
The naive greedy algorithm, which includes enumerating and storing each pair of profiles distinguished by each attribute, has a storage complexity of $O(|\mathcal{U}||B|^2)$ and computational complexity of $O(s|\mathcal{U}||B|^2)$. Instead, we design a greedy algorithm whose storage complexity is $O(|\mathcal{U}||B|)$ and its computational complexity is $O(s|\mathcal{U}||B|)$. 

Our proposal is shown in Algorithm \ref{alg:link}.  We iteratively partition the set of profiles in $B$ into anonymity sets, where all profiles in an anonymity set have identical values at all items of $K$ (i.e., they are not distinguished by any items in $K$). Our goal is to compute a partitioning where each partition (i.e., anonymity set) is as small as possible.

To do so, in each iteration, we compute a separation metric of each item for the current partitioning $C$, 
which is the total number of separated pairs of profiles over all partitions (in Line 7-11). 
For example, if a partition is split into sub-partitions $P_1$ and $P_2$ by item $i$ (where all profiles in a partition share identical value at position $i$), then the separation of $i$ with respect to $P$ is $|P_1|\times|P_2|$. The total separation of $i$ is the sum of separations over all partitions. 
Then, the item with the largest separation, denoted by $i_{\mathit{sep}}$, is selected  (Line 12). Finally, each partition  in $C$ is split by $i_{\mathit{sep}}$ into two sub-partitions (Line 17-21) such that a sub-partition contains all profiles from $C$ which have identical values at position $i_{\mathit{sep}}$. To speed up the computation of the separation metric, we pre-compute a table called $\mathit{items}$ which maps each profile $u$ to the set of items which are contained by $u$ (in Line 2 of Algorithm \ref{alg:link}).  Again, this pre-computation runs in $O(|\mathcal{U}||B|)$.

\begin{figure*}[h!]
    \centering
        \includegraphics[width=13cm]{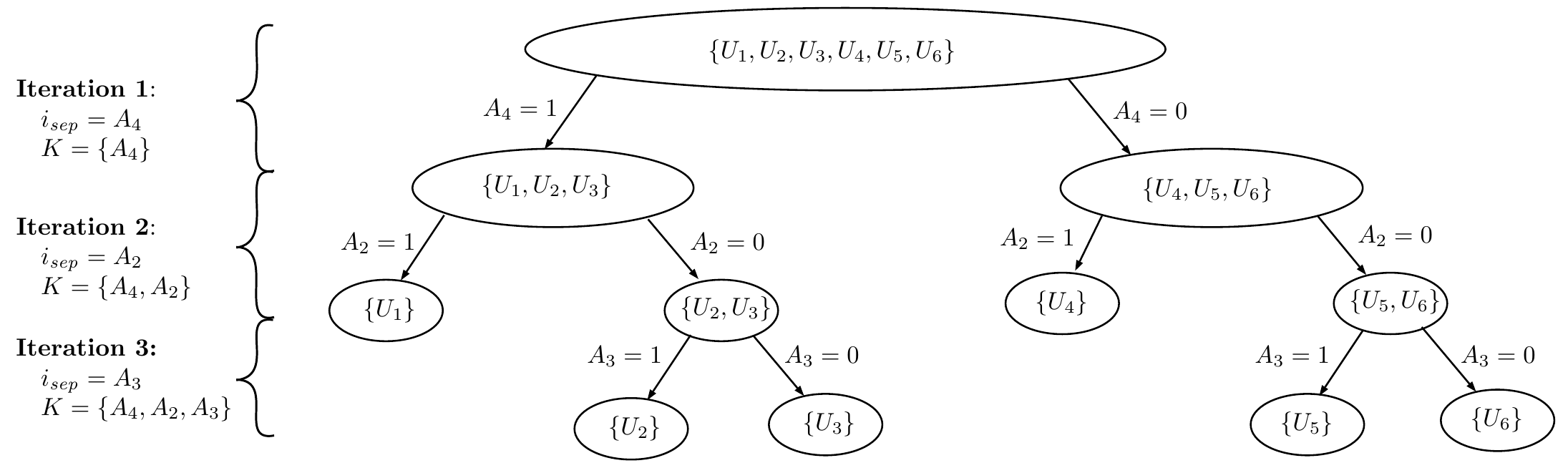}
        \caption{General fingerprinting on the dataset of Table \ref{tab:ex}: Greedy algorithm builds a binary tree of partitions, where the leaves constitute the final anonymity sets of the fingerprint $K = \{A_2, A_3, A_4\}$. \label{fig:group}}
\end{figure*}

\begin{theorem}
Suppose that $s=3$ and we want to compute the general fingerprint of Table \ref{tab:ex}. The operation of the algorithm can also be illustrated by a binary tree which is shown in Figure \ref{fig:group}. In the first iteration, we compute the separation metric of all attributes $A_1$, $A_2$, $A_3$, $A_4$ which are $5\times 1$, $4 \times 2$, $4 \times 2$, and $3 \times 3$, respectively. Hence, we select $A_4$ which separates all users into two partitions $\{U_1, U_2, U_3\}$ and $\{U_4, U_5, U_6\}$. In the second iteration, we again compute the separation metric of $A_1$, $A_2$, $A_3$ over the two partitions, which are $2 \times 1 + 0 = 2$, $2\times 1 + 2 \times 1 = 4$, $2\times 1 + 2 \times 1 =4$, respectively. Hence, we can select $A_2$ or $A_3$. We select, say, $A_2$, which produces partitions $\{U_1\}, \{U_2, U_3\}, \{U_4\}, \{U_5, U_6\}$. Next, we compute  again the separation metrics of $A_1$ and $A_3$ over the new partitions which are $0 + 1 \times 1 + 0 +0 = 1$ and $0 + 1\times 1 + 0  + 1 \times 1=2$, respectively. Therefore, we finally select $A_3$. As 3 attributes have been selected, we terminate and return $K = \{A_2, A_3, A_4\}$ as the general fingerprint.
\end{theorem}

Notice that the maximum number of partitions in any iterations is at most $|B|$ (i.e., the number of all records), and we only maintain the partitioning resulted by the last iteration.
The complexity of computing the separation metric of each attribute takes $O(|\mathcal{U}||B|)$ (Line 7-11) in each iteration, as set-memberships can be checked in $O(1)$ using hash-maps. Similarly, the separation itself (Line 17-21) runs in $O(|\mathcal{U}||B|)$, which means that the total complexity is $O(s|\mathcal{U}||B|)$ (Line 5-22).

\begin{algorithm}[ht!]
\caption{Greedy Approximation for General Fingerprinting\label{alg:link}}

\small
\DontPrintSemicolon
\KwIn{Fingerprint size $s$, Universe $\mathcal{U}$, Dataset $B$}
\KwOut{Fingerprint items $K$} 
$K := \varnothing$\;
$\mathit{items}[u] := \{i\, |\, i\in \mathcal{U} \wedge u_i = \mathsf{True}\}$ for all $u \in B$\;
$\mathit{users}[i] := \{u\, |\, u\in B \wedge u_i = \mathsf{True}\}$ for all $i \in \mathcal{U}$\;
 $C := B\qquad$ \texttt{// anonymity sets}\;
\While {$|K| < s$}
{
\texttt{// compute the separation of all items}\;
$\mathit{separation}[i] := 0$ for all $i \in \mathcal{U}$\;
\For {$S \in C$}
{
\For {$i \in \bigcup_{u \in S} \mathit{items}[u]$}
{
$t[i] := |\{u\, |\, u \in S \wedge i \in \mathit{items}[u] \}|$\;
$\mathit{separation}[i] := \mathit{separation}[i] + t[i] \cdot (|S| - t[i])$\;
}
}
$i_{\mathit{sep}} := \arg \max_{i \in \mathcal{U} \setminus K} \mathit{separation}[i]$\; 
\texttt{// if no further separation is possible}\;
\If{$\mathit{separation}[i_{\mathit{sep}}] = 0$}  
{ $\mathbf{break}$\; }
\texttt{// splitting all partitions with $i_{\mathit{sep}}$}\;
$T = \varnothing$\; 
\For {$S  \in C$}
{
$V := \{u\, |\, u \in S \wedge i_{\mathit{sep}} \in  \mathit{items}[u]\}$\;
$T:= T \cup \{\{V\}, \{S \setminus V\}\}$\;
}
$C := T$\;
$K:= K \cup \{i_{\mathit{sep}}\}$\;
}
\KwRet{$K$}
\end{algorithm}  

Finally, we note that Algorithm \ref{alg:link} can also be adapted to compute the minimum number of items which separate all users in $B$ (this is referred to as the Minimum Key Problem in \cite{masking_vldb}). To do so, we only need to set $s$
to  $|\mathcal{U}|$ in Algorithm \ref{alg:link}, which means that the running complexity becomes $O(|\mathcal{U}|^2|B|)$. By contrast, the best prior greedy solution of this problem, which is also proposed in \cite{masking_vldb}, runs in $O(|\mathcal{U}|^3|B|)$. However, this is impractical if the number of attributes/items is large. 
Although several probabilistic relaxations of this problem is considered in \cite{masking_vldb} in order to reduce running complexity, they also degrade the accuracy of the solution.

\section{Applications}
\label{sec:applications}
In this section, we describe some potential applications of targeted and general fingerprinting in practice. The first application relates to fingerprinting smartphone applications, which has also been partially studied in \cite{appuni15}. The second application shows how to fingerprint users using the list of (non-)installed fonts in order to track them while they are browsing the web. Finally, we show a use-case of mass de-anonymization in location datasets using our fingerprinting techniques, which can serve as a potential investigation tool for authorities.

\subsection{Smartphone app fingerprinting}

\begin{figure*}[h!]
    \centering
    \begin{subfigure}[b]{0.48\textwidth}
	    \centering
        \includegraphics[width=0.65\textwidth]{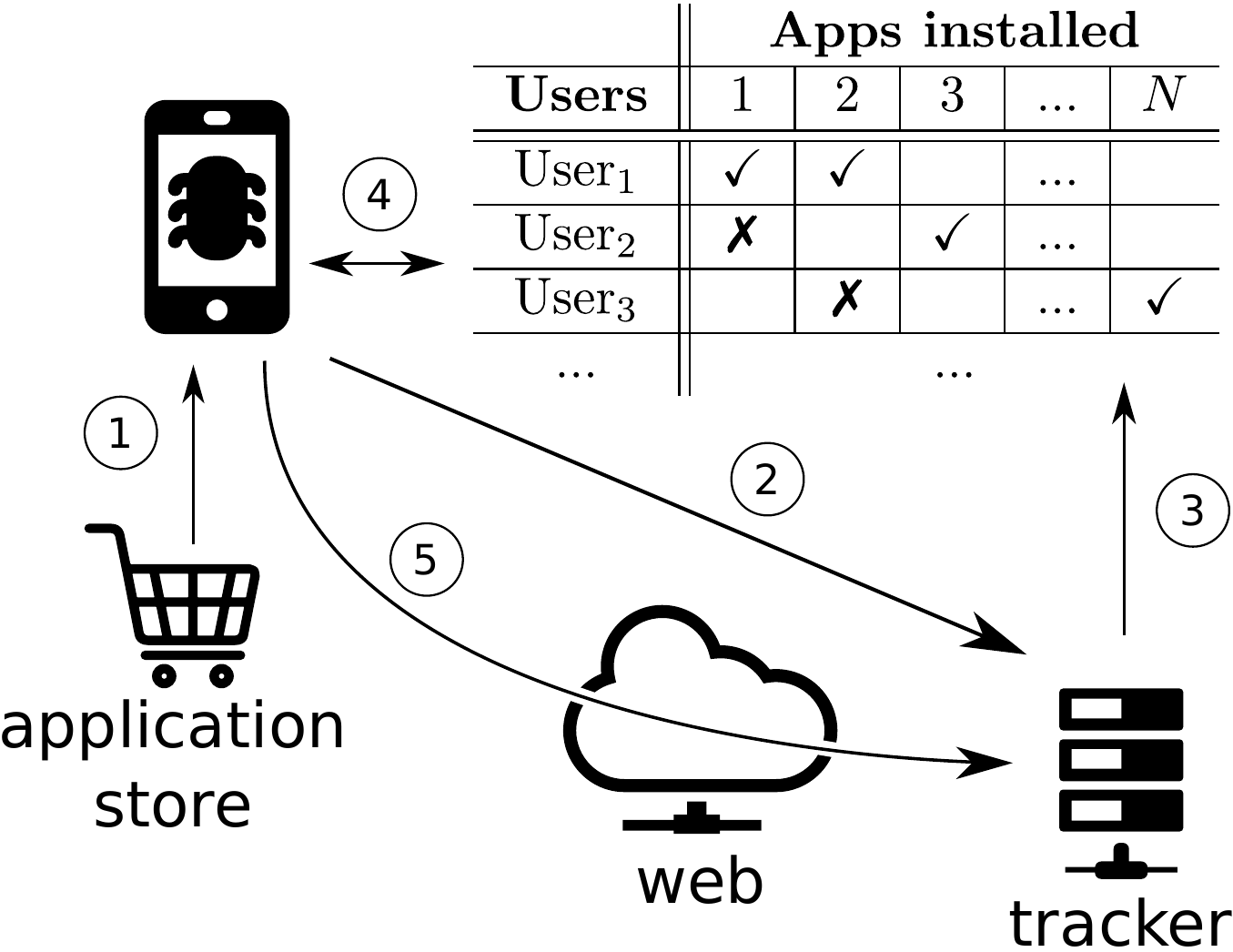}
        \caption{Re-identification: after the tracker app is installed \circled{1}, the individual fingerprint apps of the targeted users, whose identities are known, are retrieved from the 3rd party tracker \circled{2}-\circled{3}. These fingerprints are matched with the list of (non)-installed apps on the device \circled{4}. If there is a positive match, the browsing activities of the user along with his identity is tracked using a cookie created by the tracker app \circled{5}.}
        \label{fig:intro:deanon}
    \end{subfigure}
    ~
    \begin{subfigure}[b]{0.48\textwidth}
	    	\centering
        \includegraphics[width=0.6\textwidth]{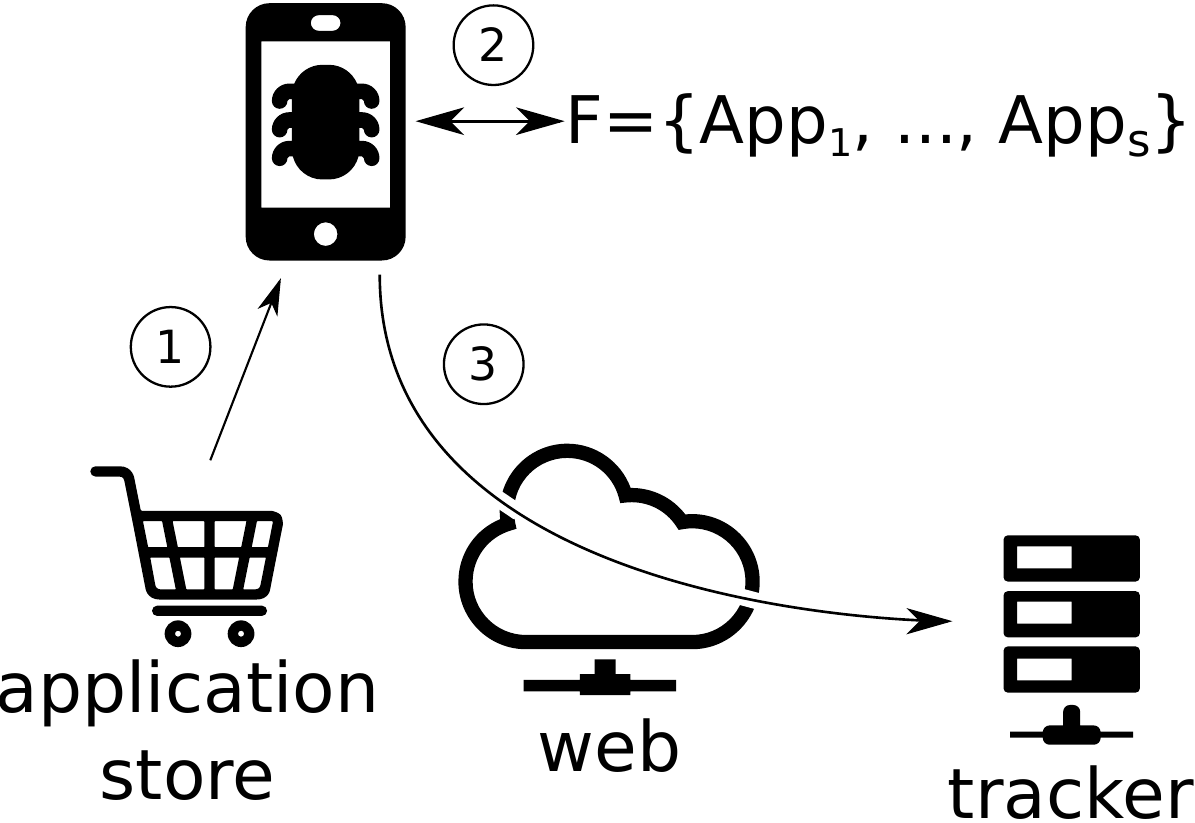}
        \caption{Linking attack: the tracker app is installed with the general fingerprint apps which distinguish all users the most \circled{1}. After the fingerprint value is computed by checking the existence of the fingerprint apps  on the device \circled{2}, a cookie is created by the tracker which is bound to this fingerprint value. As a result, the browsing activities of any user in the group can be tracked \circled{3}.\newline}
        \label{fig:intro:linking}
    \end{subfigure}
    \caption{Targeted fingerprinting (for re-identification) and general fingerprinting (for linking) using the list of installed apps on a smartphone.}\label{fig:intro}
\end{figure*}

It has been shown that knowing the presence of four randomly chosen applications on a device are enough to re-identify users in a dataset of 54,893 users \cite{appuni15}. Moreover, installed apps also reveal a lot about the owner's interests, and Twitter also used this information for targeted interest-based advertising among others \cite{twitter14}. In this section, we show how a third party can use such 
information for tracking. The main idea is to install a malicious application on the user's device which executes the fingerprint queries and identifies the fingerprint value of the user. As some operating systems, such as iOS, limit the number of apps whose existence can be checked by another app, the queries should be optimized such that their result distinguish as many users as possible.
In particular, we envision a re-identification scenario using targeted fingerprinting (see Fig. \ref{fig:intro:deanon}), when the browsing activities of some targeted users are tracked, and also a linking scenario using general fingerprinting (see Fig. \ref{fig:intro:linking}), when the browsing activities of multiple users are tracked without re-identifying any of them.

In both cases, when the (individual or general) fingerprint apps are provided to the malicious app, it checks their existence on the device (i.e., constructs the fingerprint value). Then, it opens a legitimate URL in the browser which redirects to a third party tracker in order to generate and store a tracking cookie of the tracker as a visited first party in the browser\footnote{In order to circumvent the third-party blocking of Safari, the legitimate URL specifies a malicious website which redirects to the third-party tracker. The tracker then becomes a first party and can set a tracking cookie in the browser. Finally, the tracker redirects to the initial website, and the device can be tracked across websites.}. This cookie contains the user/device identifier which is the fingerprint value. When the same browser visits a site having this tracker, the user will be tracked using the previously stored cookie (and also re-identified in case of targeted fingerprinting).

We must note that, in contrast to traditional web cookies, the identifier stored in the cookie is strictly bound to a persistent, behavioural attribute (i.e., the list of installed/non-installed apps). This means that even if the user deletes the cookie, he can still be re-linked to the same identifier by re-generating the same cookie  based on possibly the same  list of (non-)installed apps. The identifier even persists, if the application is removed and then re-installed, or other applications are installed that can generate the same identifier.

Next, we provide the details of the attack, which consists of three phases: (1) \emph{Identify the fingerprint apps (queries) of the targeted user (targeted fingerprinting) or a group of users (general fingerprinting)}, (2) \emph{Retrieving the list of fingerprint apps on users' device}, (3) \emph{De-anonymize the targeted user and/or track user activities}. As the third phase has been detailed before, we will focus on the first two phases.

\subsubsection{Identifying the fingerprint apps} For targeted fingerprinting, the malicious 3rd party tracker, called as adversary henceforth, first needs to obtain the list of (non-)installed apps of a known, targeted user. This is relatively easy by consulting with some social media where users often post the applications they use, or simply retrieving the app list directly from the user's device (if the user has iOS 9, the adversary may get access to the full list of installed apps before the user upgrades to iOS 9, see below for details). As another work suggests \cite{appuni15}, the adversary does not need to get the complete list of applications, but only 4 random apps are sufficient to re-identify the user with 95\% of the times. We show later that, by using the greedy optimization algorithm, described in Section \ref{sec:ind_algo}, the adversary can re-identify most users with even fewer queries.

For both targeted and general fingerprinting, the adversary needs to have access to a subset of user profiles, denoted by  $B$ in Algorithms \ref{alg:deanon} and \ref{alg:link}, in order to identify the fingerprint apps of the users to be tracked. In particular, for targeted fingerprinting, the adversary identifies at most $s$ apps, using Algorithm \ref{alg:deanon}, whose existence need to be checked on the targeted user's device to construct his/her individual fingerprint. 
As we detail below, the value of $s$ changes depending on which operating system the app is built for. In certain cases, iOS allows to query at most 50 apps, while in other cases the queries need to be pre-defined and cannot be changed later when the app is installed, which is reminiscent of our general fingerprinting problem.

\begin{figure}
	\centering
	\includegraphics[width=0.48\textwidth]{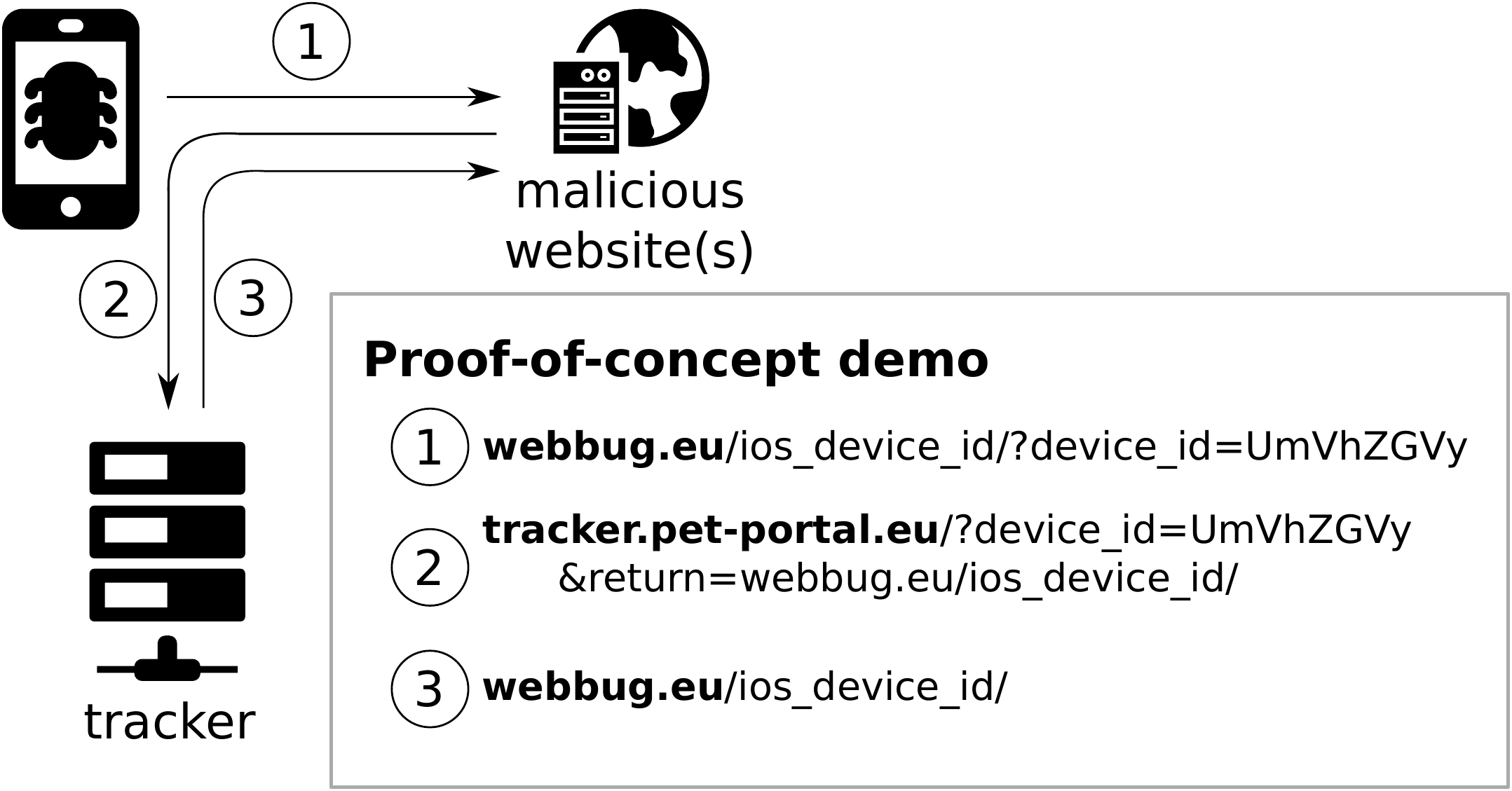}
	\caption{Example on how the device identifier could be leaked to the web. The malicious application opens a URL that seems to be legitimate, but it contains the hidden or obfuscated device identifier \circled{1}. In order to circumvent the potential third-party blocking of Safari, the malicious websites redirects to a third-party tracker \circled{2}. The tracker then is now becomes a first party and can set a tracking cookie. Finally, the tracker redirects to the initial website \circled{3}, while it also remains able to track the devices across websites.}
	\label{fig:ios_leak}
\end{figure}

For general fingerprinting, the dataset is used to 
identify at most $s$ fingerprint apps, using Algorithm \ref{alg:link}, whose existence need to be checked on \emph{every} group member's device to construct a fingerprint for each group member which sufficiently distinguish them. This dataset can contain the profiles of the group themselves, or another subset of the \emph{same} population. After calculating the user/device identifier, this could be used as an identifier for web tracking; we show how in Fig. \ref{fig:ios_leak} (with a proof-of-concept demo).

\subsubsection{Retrieving the fingerprint apps on iOS} 
\label{sec:ios}
Equipped with the list of the fingerprint apps, the malicious application needs to check their existence on the device.
To the best of our knowledge, Android allows to retrieve the complete list of installed apps, but Apple has introduced some limitations with the release of iOS 9 \cite{appleios9}.
On iOS, applications can use the \texttt{canOpenURL()} call to determine if another application is available on the system (i.e., whether it can be called).  That is, each fingerprint app query corresponds to a call of \texttt{canOpenURL()} specifying a URL. The call returns a boolean value indicating whether or not the specified URL's scheme can be handled by some app installed on the device.
However, Apple introduced two restrictions that apply when the program is compiled on iOS 9 \cite{appleios9}. Applications built for systems before iOS 9 but executed on iOS 9 can invoke \texttt{canOpenURL()} with at most 50 distinct apps. This number is reset each time the user re-installs or upgrades the app. Applications built for iOS 9 and also executed on iOS 9 have to declare in advance which other applications they would like to check with \texttt{canOpenURL()}, making abuses noticeable well in advance. 
Every \texttt{canOpenURL()} call which is over the limit of 50, or specifies a non-declared app, will return a negative response.

We emphasize again that these limitations are in effect only when a user builds an app on iOS 9, leaving users of older systems vulnerable. The proportion of such systems was measured to be $23\%$ according to a statistic revealed by the Apple Store (as of February 22, 2016) \footnote{Statistic published by the Apple Store: \url{https://developer.apple.com/support/app-store/}}.

\subsection{Font fingerprinting}
\label{sec:font_app}
Fingerprinting projects, such as the Panopticlick\footnote{Website: \url{https://panopticlick.eff.org}} \cite{pan10} and the cross-browser fingerprinting project\footnote{Website: \url{https://pet-portal.eu/fingerprint/}; newer version available at: \url{https://fingerprint.pet-portal.eu}} \cite{cross12} have revealed that websites can effectively query browser properties to track their visitors. In particular, it has been shown that detecting fonts is an important technique to track users as their entropy is one of the highest among the available identifiers \cite{pan10}, and they can be queried independently from the browser \cite{cross12}. Font detection mechanisms were also shown to be used in device tracking mechanisms \cite{cookieless13}.

\begin{figure*}
	\centering
	\includegraphics[width=0.6\textwidth]{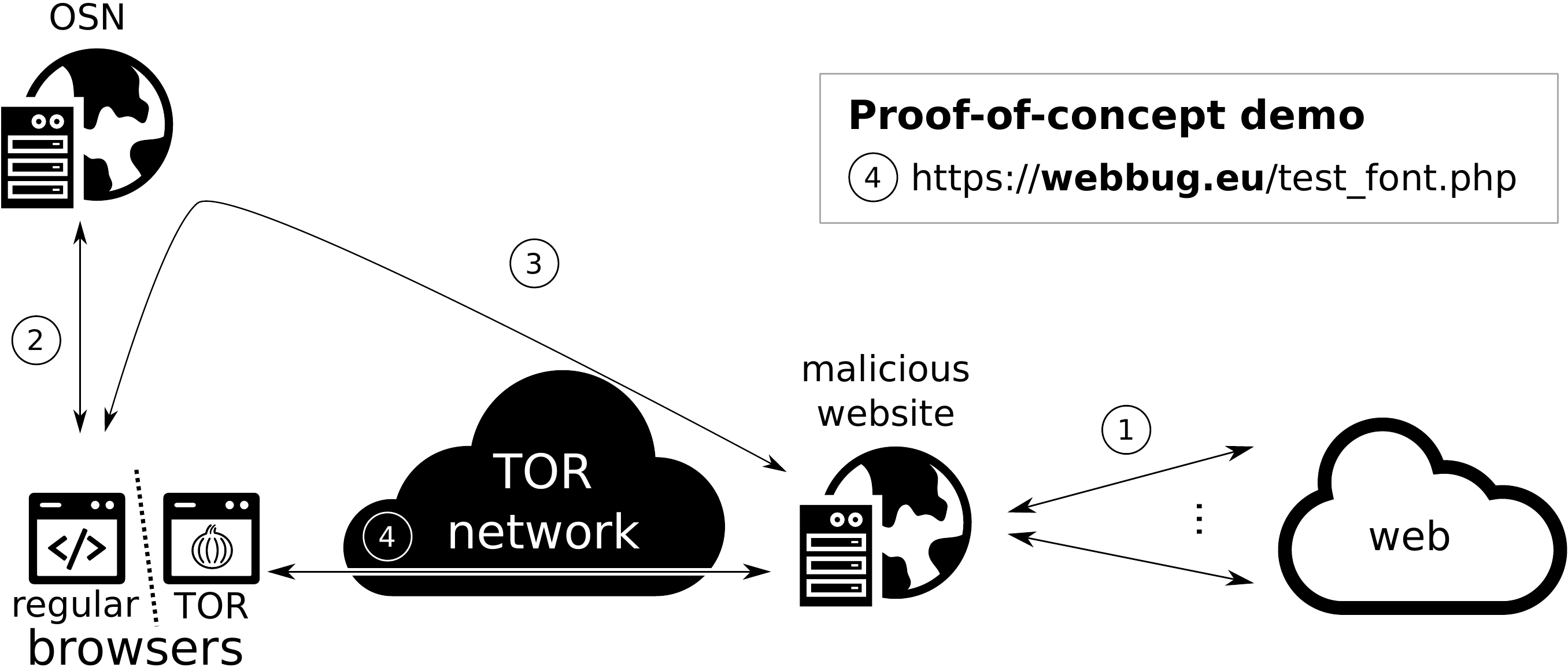}
	\caption{Attacks against TOR browser. Initially, the malicious websites (or collaborating sites) build a large dataset of (font) profiles by installing trackers (or web bugs) on websites which are also accessed without TOR \circled{1}. In case of \emph{targeted fingerprinting}, when the user visits a site containing a malicious tracker \circled{2}, the adversary gets informed by its trackers and retrieves the complete profile of the targeted user \circled{3}. Then, using the collected dataset of font profiles, the adversary can compute the fingerprint fonts of the user, which can be later queried in order to de-anonymize the user when he visits a site through TOR containing a malicious tracker \circled{4}. In case of \emph{general fingerprinting}, the adversary computes the general fingerprint using the  collected dataset of font profiles,  and queries these fonts from each TOR visitor \circled{4}.}
	\label{fig:tor_attacks}
\end{figure*}

One cutting edge web browser that promises anonymity is called the Tor Browser Bundle \footnote{The TOR project website can be found here: \url{https://www.torproject.org}} (TBB) or the Tor Browser, whose anonymity guarantees have been widely studied \cite{Cai:2012, Danezis:2015}. Up to version 5.5, TBB developers tackled font-based tracking by introducing two limits: one on the number of font load trials, and one on the number of successful font loads \cite{tord15}. These parameters are called \texttt{browser.display.max\_font\_count} and \texttt{browser.display.max\_font\_attempts}, and by default they are set to a value of 10. These options can be changed by accessing \texttt{about:config}. Detecting the existence of fonts on a system is relatively easy using Javascript, which can be embedded on the visited website: it can be measured if the rendered text has a different width and height compared to the system default \footnote{Example: \url{http://www.lalit.org/lab/javascript-css-font-detect/}}. Notice that TBB allows the execution of Javascripts by default due to their prevalence on the web today. TBB addresses two types of attacks, among others, which we analyze in this paper: de-anonymization (see Section 3.1.2 in \cite{tord15}) and linking attacks (see Section 3.1.4 in \cite{tord15}). Our attack schemes are depicted in Fig. \ref{fig:tor_attacks}.

\emph{De-anonymization} occurs when an attack correlates TBB and non-TBB activities of a targeted user. First, the adversary collects the available fonts on the user's system when the user visits a website with a regular, non-TBB browser, thus the list of fonts can be associated with the user's identity. We note that this does not mean that the site is under a total control of the adversary. This can also be a site where a malicious tracker (or web bug) is installed. Then, the adversary selects at most 10 fingerprint fonts whose existence need to be checked on the user's system to re-identify him/her.  When the user visits this or another malicious website via Tor Browser, this site queries these fingerprint fonts through Javascript. As a result, ``private'' (TBB) and ``public'' (non-TBB) activities of the web user can be linked. De-anonymization is analogous to our targeted fingerprinting problem, where the maximum number of fingerprint queries is limited to 10.

\emph{Linking attack} refers to linking the activities of a user together within TBB. As opposed to de-anonymization, the identity of the user is not known (i.e., the attack does not require a visit to a site with a regular browser), and the goal is to track the user using TBB. In this case, the adversary needs to identify at most 10 fingerprint fonts which can distinguish \emph{any} TBB user from the rest. This means that a malicious tracker queries the \emph{same} set of 10 fonts of each TBB user visiting a site, and the results of these queries serve as a fingerprint for each user. 
This is  analogous to our general fingerprinting problem, where the maximum number of fingerprint queries is limited to 10.
Notice that while these fingerprints may not be sufficiently unique to every user in the whole TBB community, they should be quite unique per site. According to a measurement based on TBB usage data of 2012-2013 \cite{torstats14}, the average number of daily Tor Browser users per country was well below $1\%$, e.g., $0.08\%$ in Europe, meaning that even visitors of well known websites receiving a large daily traffic could be easily identified even if they would use TBB. 

In order to identify the fingerprint fonts, the adversary needs to have access to a subset of all profiles in the TBB population, where a profile contains the list of installed and non-installed fonts of a user.

\textbf{Note on deprecation:} In the beginning of our investigation, we found out that the font limitation was not working at all in the Tor Browser, making TBB users vulnerable to the above attacks without limitation (the current stable version was TBB 5.0.4 at that time).
We published a blog post to demonstrate the feasibility of the above attacks \cite{tortrack15}, and informed the developers about this issue. It turned out that developers had already been working on other workarounds to defeat font fingerprinting \cite{torframe13}, and the lack of the above limitation on the number of font queries was due to implementation difficulties in newer Firefox versions. The new countermeasure, which is shipping the Tor Browser with a pre-defined set of fonts that the browser can only use, was introduced two months later with TBB 5.5 at January 26, 2016. While this makes our fingerprinting techniques deprecated in the context of TBB, we believe that our work can be a useful warning for developers who may use similar countermeasures in the future, as the case of application fingerprinting also shows.

\subsection{Mass de-anonymization in location datasets}
\label{sec:app_location}
We can also use fingerprinting techniques in a quite different context than online tracking, that is, to de-anonymize a location database (e.g., a Call Detail Record dataset). Suppose that the adversary has access to a weakly anonymized location dataset containing the set of visited locations per individual in a city (time information is omitted for the sake of simplicity). For example, this dataset contains pseudonymized records (i.e., only direct personal identifiers are removed), but location information are kept intact. The goal of the adversary is to re-identify as many users as possible in this dataset, i.e., to perform a \emph{mass de-anonymization attack} \emph{without knowing any locations of the victims a priori}\footnote{Otherwise, the adversary has good chance to locate the victim's record based on \cite{Nature13}}, but only their personal photos (e.g., by crawling a social media of the city). 
The plan of the adversary is to mount at most $s$ cameras at different spots of the city, and re-identify all individuals observed by any of these cameras by using some face recognition mechanism and the photos that she has. Knowing the fact that an individual does (not) visit these camera spots (i.e., the visiting pattern of individuals with respect to only these spots), the adversary locates the record  which has the observed visiting pattern in the anonymized dataset, and learn all other unknown visited places of the individual.\footnote{It is not reasonable to assume that
the adversary can mount any number of cameras due to budget constraints.}  

In order to maximize the number of de-anonymized users in the dataset, the adversary wants to identify at most $s$ camera spots which can distinguish \emph{all users} in the anonymized dataset as much as possible. In particular, the values of \emph{any} record at these spots should be unique in the dataset, or at least the number of records sharing the same values at these spots should be minimized.
This goal is reminiscent of our \emph{general fingerprinting problem}. Hence, the adversary runs our greedy heuristic in Algorithm \ref{alg:link}, with input $s$ and the anonymized dataset, in order to identify these fingerprint spots. 

Although deploying cameras at any spot of a city is unlawful and easily detectable, we believe that the above scenario is not far-fetched at all. Authorities, who have the right to install or access such surveillance cameras, may be interested in the above ``privacy-friendly'' approach of  learning the trajectories of suspects. Indeed, authorities do not need to ask telecom companies to reveal the exact identity of any of their customers, but only their anonymized CDRs.

\section{Experimental results}
In this section, we compute the uniqueness of targeted and general fingerprints on real-world datasets. In particular, we consider three real-world datasets, each corresponding to an application described in Section \ref{sec:applications}, which can be used by the adversary to identify the fingerprint items of users. We report the size of the anonymity sets of all targeted and general fingerprints for each dataset.

\subsection{Datasets and ethical considerations}

\subsubsection{Smartphone applications}
\label{sec:app_dataset}

\paragraph{Dataset}  
The dataset comes from the Carat research project \cite{Oliner:2013}.
The dataset includes data from $54,893$ Carat Android users between 11/03/2013 and 15/10/2013  \cite{Truong:2014}.
During this period, the Carat app\footnote{\url{http://carat.cs.helsinki.fi}} was collecting the list of running apps (and not the list of all installed apps) on users' devices when the battery level changes.
As collecting the list of running apps multiple times over more than 7 months is likely to sum up to the set of all installed apps of a user, we consider a record as the set of installed applications in this paper, even if a record might not be the complete set of installed apps all the time.

We removed system apps from all records because they are common to all users.
Without system apps, our analyzed dataset contains $92,210$ different applications whereas the total number of apps available on the GooglePlay were around 1 million during this time\footnote{\url{http://en.wikipedia.org/wiki/Google_Play}}.
Table \ref{tab:D} summarizes the main characteristics of our dataset.

\paragraph{Ethical considerations}
This dataset contains the list of installed smartphone applications per user.
The data were collected with the users' consent, and they were explicitly informed that their data could be used and shared for various research projects.
In fact, the Carat privacy policy (available at \url{http://carat.cs.helsinki.fi}) clearly specifies that ``Carat is a research project, so we reserve the right to publish our results online and in academic publications. 
We also reserve the right to release the data sets into the public domain.''
Also, the dataset was shared with us by the Carat team in a pseudo-anonymised form. 
In particular, identifiers were removed, and each application name was replaced with its SHA1 hash. 
It contained $54,893$ records~\cite{Truong:2014}, i.e. one record per user. 
Each record is composed of the list of applications installed and run by the user.
Furthermore, the data sharing agreement that we signed, stipulated that we cannot use the dataset to deanonymize the users in the dataset.

\begin{table}
\begin{minipage}{.5\textwidth}
\centering
	\begin{tabular}{|l|l|}
	\hline
	\# of records & $54,893$ \\ \hline 
	\# of all apps in the dataset & $92,210$ \\ \hline
	Maximum record size  & $541$\\ \hline
	Minimum record size  & $1$ \\ \hline
	Average record size &  $42$ \\ \hline
	Std.dev of record size & $39$ \\ \hline	
	\end{tabular}
	\caption{\label{tab:D} Characteristics of the smartphone app dataset}
\end{minipage}
\begin{minipage}{.5\textwidth}
\centering
	\begin{tabular}{|l|l|}
	\hline
	\# of records & $43,656$ \\ \hline 
	\# of all fonts in the dataset & $81,195$ \\ \hline
	Maximum record size & $6,072$ \\ \hline
	Minimum record size & $1$ \\ \hline
	Average record size & $143$ \\ \hline
	Std.dev of record size & $118$ \\ \hline
	\end{tabular}
	\caption{\label{tab:font_dataset} Characteristics of the font fingerprint dataset}
\end{minipage}
\end{table}

\subsubsection{Fonts used by web browsers}
\label{sec:font_dataset}

\begin{figure}[h!]
    \centering
    \begin{subfigure}[b]{0.48\textwidth}
    		\centering
		\begin{tabular}{l|c|c}
			~                 & \rot{Panopticlick} & \rot{current}\\ \hline
			User agent string & $10.0$               & $7.18$ \\ \hline
			Timezone          & $3.04$               & $2.23$ \\ \hline
			All fonts         & $13.9$               & $7.79$ \\ \hline
			Plugins           & $15.4$               & $7.91$ \\ \hline
			Screen            & $4.83$               & $3.34$ \\
		\end{tabular}
        \caption{Comparison of entropy values for each attribute.}
        \label{fig:tor:entropy}
    \end{subfigure}
    ~
    \begin{subfigure}[b]{0.4\textwidth}
        \includegraphics[width=\textwidth]{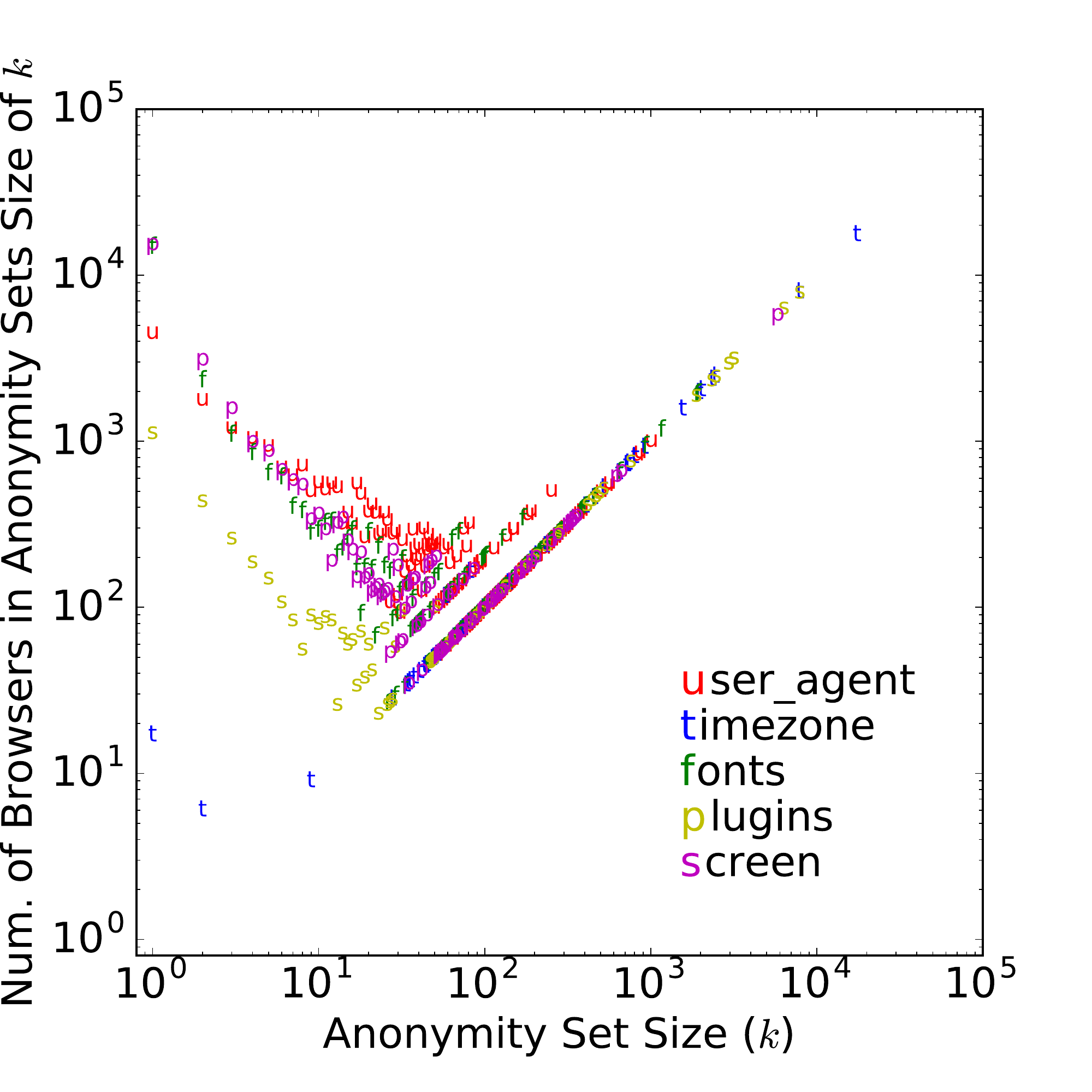}
        \caption{Number of users w.r.t. the sizes of the anonymity set they belong to.}
        \label{fig:tor:as_distrs}
    \end{subfigure}
    \caption{Descriptive attributes of our datasets that also enables the comparison with previous work in \cite{pan10}.}
    \label{fig:tor}
\end{figure}

\paragraph{Dataset} 
This dataset contains the list of installed fonts, the screen size and available screen size per user.
The data collection was done on \url{fingerprint.pet-portal.eu} where users could test and compare their OS fingerprint in different browsers. Data was collected with the purpose of analyzing OS fingerprint uniqueness. In particular, users were advised to run the test in different browsers which collected the list of installed fonts, among others, and compare the calculated fingerprint if they match
The website have collected more than $107,102$ font profiles between 05/04/2012 and 22/09/2015, which do not necessarily belong to distinct users due to the purpose of the project. In our experiments, we only used a subset of users and attributes that were collected, and the data was cleaned as follows.

In order to remove possible duplicates (a user could visit the fingerprinting site multiple times), several records were removed by comparing hashed IPs and stored usernames. If username was not given by a person who conducted the test, the site stored the font profile of the user in an evercookie\footnote{Evercookies: \url{http://samy.pl/evercookie/}}. Given a list of profiles having the same IP address where creation dates were closer than $10$, only the first entry was preserved. All record duplicates having the same username were removed regardless of date differences.

The resulted database had $43,656$ records, which contained the list of detected fonts, and the screen and available screen sizes\footnote{Referring to JavaScript properties \texttt{screen.width} \& \texttt{screen.height} and \texttt{screen.availWidth} \& \texttt{screen.availHeight}} of each user. As this is the first paper using the given dataset, we provide descriptive metrics of the data in order to help understanding our results. We also provide attributes of the original Panopticlick dataset for comparison, and Fig. \ref{fig:tor:as_distrs} is also easily comparable with Fig. 3 in \cite{pan10}. This figure confirms that our dataset has been cleaned properly. The main properties of the dataset is described in Table \ref{tab:font_dataset}.

\paragraph{Ethical considerations} The data were collected with users' consent on a website that raises awareness on tracking using OS (or device) fingerprinting. On the website it is clearly described that all collected information is treated confidentially, and the goal of the data collection is to analyze uniqueness of generated OS fingerprints. Types of data that are collected is also clearly communicated, from which we used screen size, available screen size, and available fonts per user.
The data we used was shared with us with the operators of the fingerprint website, and our agreement allowed us to conduct the uniqueness study we present in this paper.

\subsubsection{Location dataset}
\label{sec:location_dataset}

\paragraph{Dataset} 
This dataset contains the list of visited locations per user in a large European city.
We use a CDR (Call Detail Record) dataset provided by a cell phone operator in Europe. A cell tower  is visited by an individual, if the operator has a recorded event at the tower related to the individual over the observed period (01/09/2007 - 17/10/2007). An event can be an incoming/outgoing call or message to/from the individual.   The dataset contains the events of $4,427,486$ users at $1,303$ towers within the administrative region of the city, where the GPS coordinates of all the towers are also available.  
In our dataset, a record contains \emph{only} the set of towers that are visited by a user over six weeks, i.e., it does \emph{not} contain the time of visits.

The average number of individuals per tower over this period was $38,817$ with a standard deviation of $50,911$.
The total area of the city which is covered by cell towers is $128.1$ km\textsuperscript{2}.
The main characteristics of our dataset and the cell towers are shown in Table \ref{tab:location}.

\begin{table}[t]
\begin{minipage}{.5\textwidth}
\centering
\begin{tabular}{|l|l|}
\hline
Dataset size & $4,427,486$ \\ \hline 
\# of all cell towers & $1303$ \\ \hline
Maximum record size  & $422$ \\ \hline
Minimum record size  & $1$ \\ \hline
Average record size & $11.42$  \\ \hline
Std.dev of record size &  $17.23$ \\ \hline
Total area of all cells &  $128.1$ km\textsuperscript{2} \\ \hline
\end{tabular}
\end{minipage}
\begin{minipage}{.5\textwidth}
\centering
\includegraphics[width=7cm]{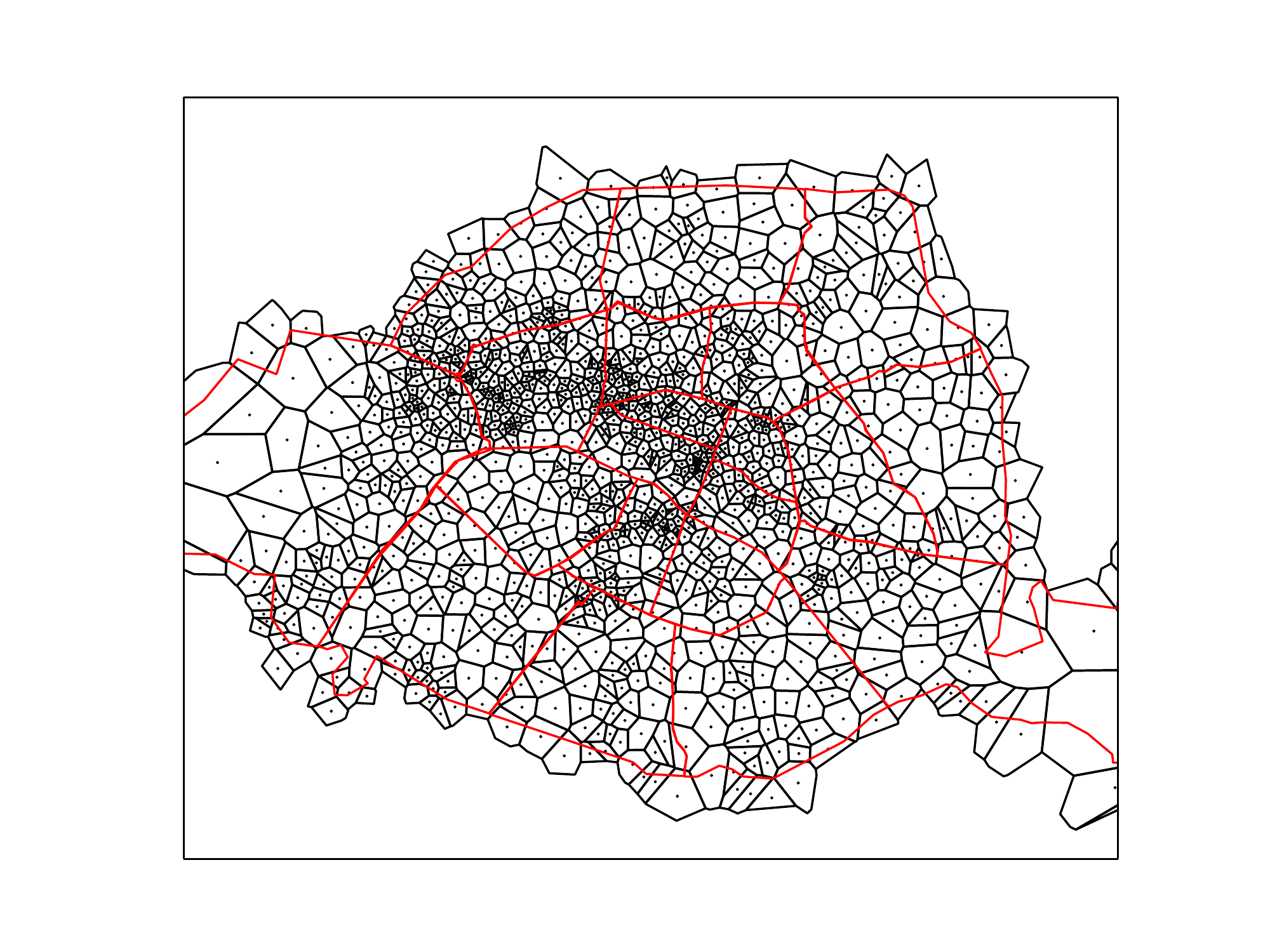}
\end{minipage}

\caption{\label{tab:location}  Characteristics of our location dataset (up) and Voronoi-tessellation of cell towers (down). Red lines denote the boundaries of districts.}
\end{table}

\paragraph{Ethical considerations}
The dataset was shared with us by the telecom operator in a pseudo-anonymised form. 
Each record in this dataset represented an event belonging to a single user with a pseudonymized phone number, i.e., the last 8 digits of the  original phone number were changed to a randomly chosen 8 digit number. 
The record included the GPS location of the cell tower, the type of the event (incoming/outgoing SMS or call) and the timestamp of the event.
Records (events) belonging to the same individual contained identical pseudonymized phone number, hence all events of an individual could be linked together.
We signed a data sharing agreement with the operator which stipulated that we cannot use the dataset to deanonymize the users in the dataset.

\subsection{Targeted fingerprinting}

\subsubsection{Smartphone users by applications}
\label{sec:ind_app}
\begin{figure*}
\begin{minipage}{.5\textwidth}
\centering
	\centering
	\includegraphics[width=.9\textwidth]{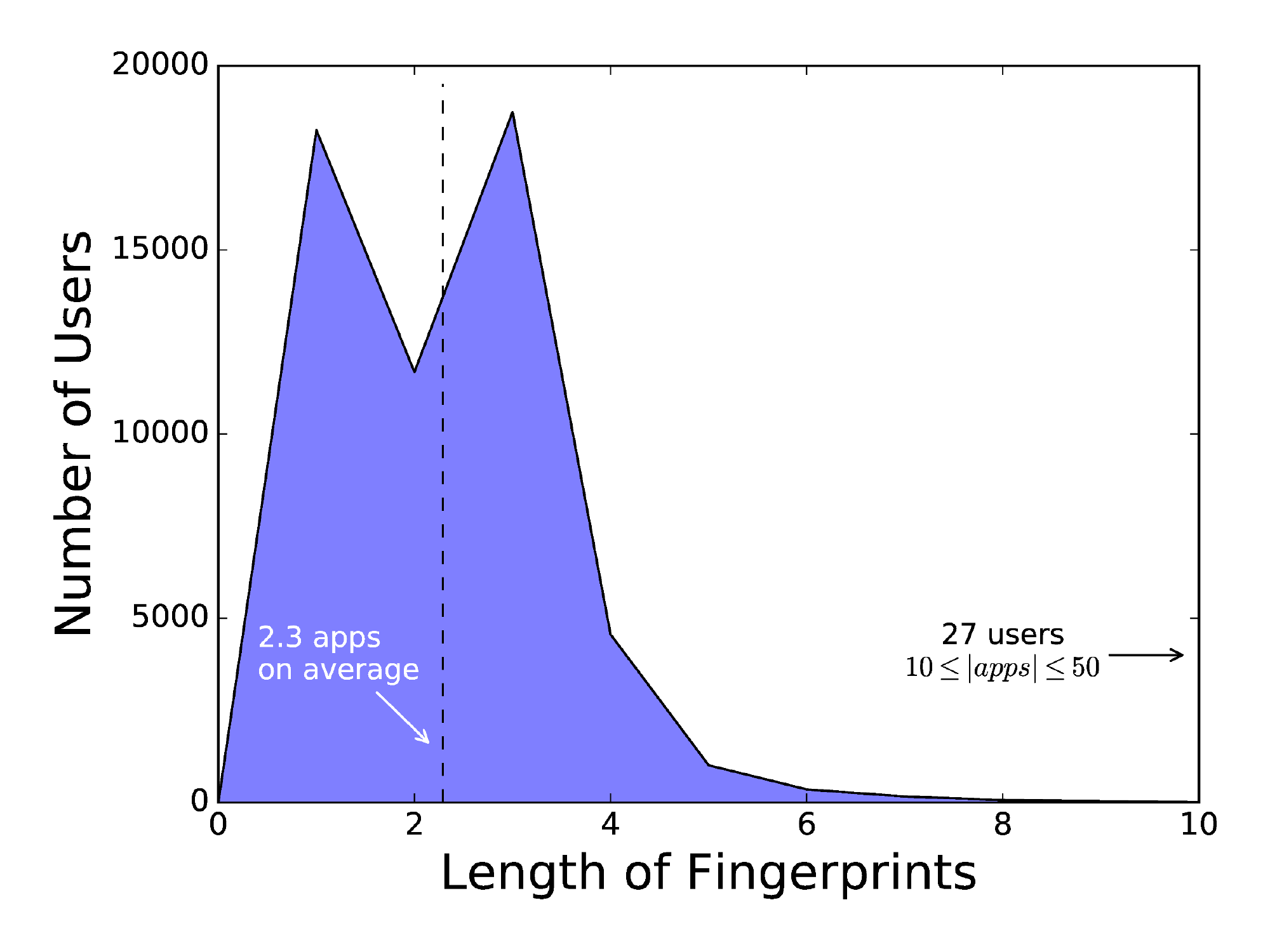}
	\caption{Distribution of the length of individual fingerprints for the app dataset. Almost all users have fingerprint size smaller than $10$, with only 27 exceptions. }
	\label{fig:app_lengths}
\end{minipage}
\begin{minipage}{.5\textwidth}
	\centering
	\includegraphics[width=.9\textwidth]{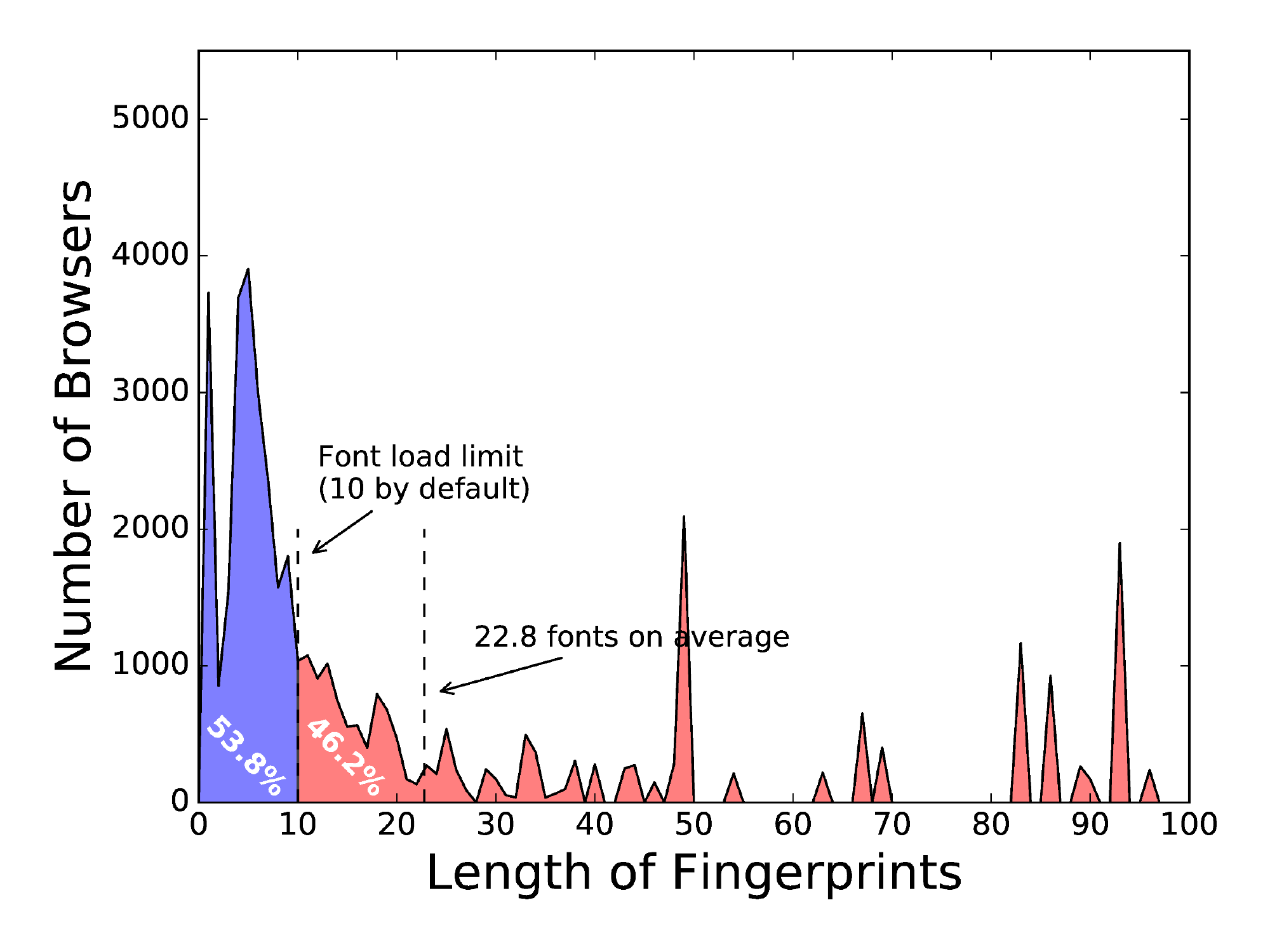}
	\caption{Distribution of the length of individual fingerprints for the font dataset. $53.8\%$ of users has a fingerprint smaller than the constraint 10. }
	\label{fig:font_lengths}
\end{minipage}
\end{figure*}

We run Algorithm \ref{alg:deanon} on the smartphone application dataset, described in Section \ref{sec:app_dataset}, to generate an individual fingerprint for each user in our dataset with different constraints on the maximum size of the fingerprint. When $s=50$ (this threshold comes from the iOS countermeasure detailed in Section \ref{sec:ios}), on average, only $2.3$ app queries per user provided the fingerprint value. The exact distribution of fingerprint length is shown in Fig. \ref{fig:app_lengths}. Our greedy algorithm generated fingerprints shorter than $50$ apps for all users. Moreover, with the exception of $27$ users,  the generated fingerprints consist of at most $10$ apps. The reason is that our algorithm can terminate before the fingerprint size reaches $s$ (see Line 7 in Algorithm \ref{alg:deanon}). We note that a user who has a  fingerprint shorter than $s$ is not necessarily unique with this fingerprint in the dataset. For example, a user may have fewer than $s$ apps installed on his phone which are also installed by other users in the dataset.

\begin{figure}[h!]
    \centering
    \begin{subfigure}[b]{0.3\textwidth}
        \includegraphics[width=\textwidth]{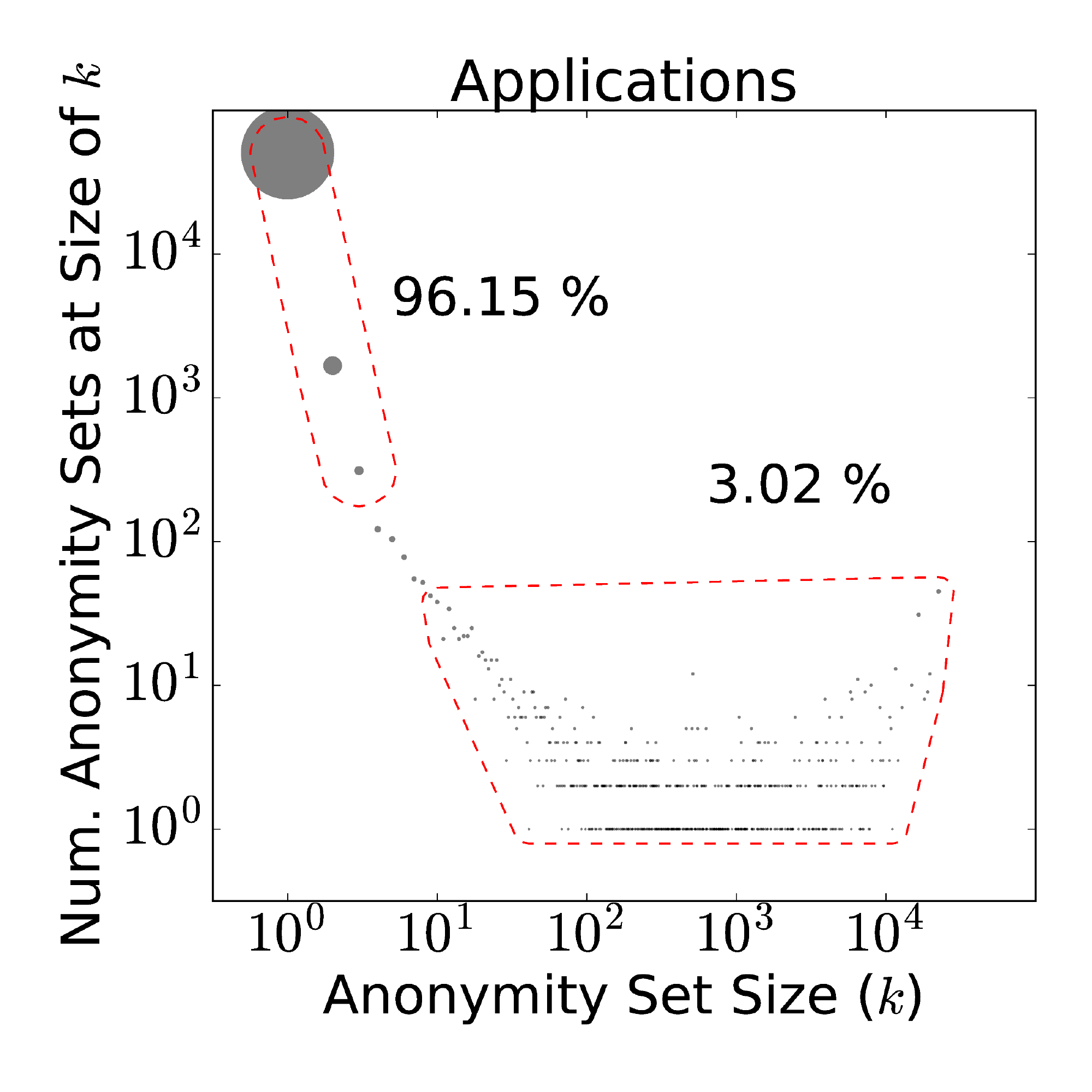}
        \caption{$s = 50$}
        \label{fig:fapps_kmap:limit50}
    \end{subfigure}
    ~
    \begin{subfigure}[b]{0.3\textwidth}
        \includegraphics[width=\textwidth]{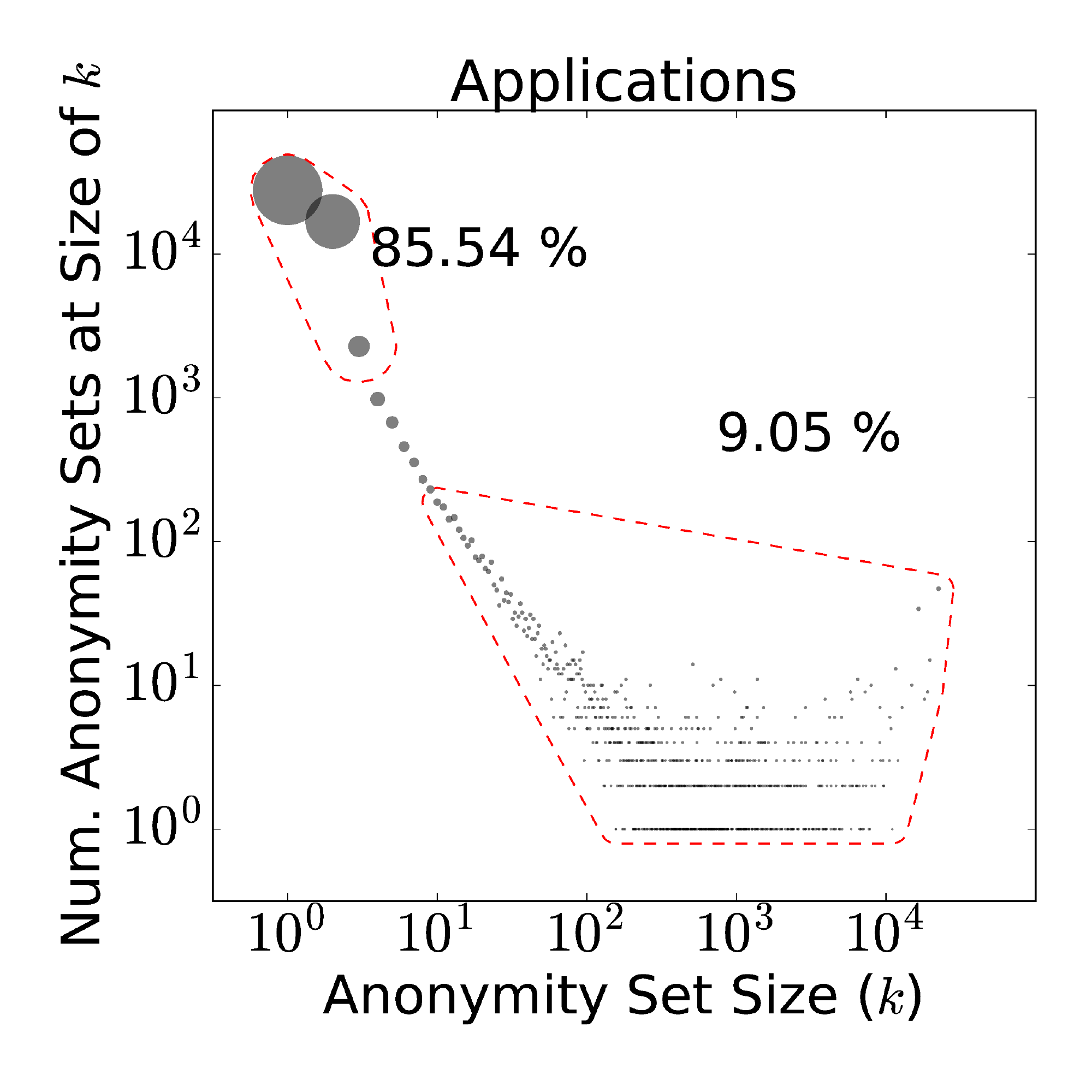}
        \caption{$s = 2$}
        \label{fig:fapps_kmap:limit2}
    \end{subfigure}
    \caption{Targeted fingerprint attacks with the app detection limit defined by Apple (a), and decreased further to only 2 applications (b).  (Note: dot sizes is proportional to the proportion of users having a particular anonymity set size.)}
    \label{fig:apps_kmap}
\end{figure}

For $s=50$, we find 4,098 users ($7.47\%$) in our dataset who cannot be uniquely re-identified and therefore belong to some anonymity sets with size greater than 1. We visualized the sizes of anonymity sets and their population in Fig. \ref{fig:fapps_kmap:limit50}. This figure shows that $96.15\%$ of users are almost unique (the size of their anonymity sets is at most 3), while only $3.02\%$ of users fall into anonymity sets with size at least 10. (Note: large anonymity sets concerning only a couple of users are possible as these anonymity sets are not disjoint.) 

We also measured the size anonymity of sets when $s = 2$ (instead of $50$); results are reported in Fig. \ref{fig:fapps_kmap:limit2}. The proportion of almost unique users decreased to $85.54\%$, eventually showing that the current privacy protection method on iOS is ineffective against fingerprinting users. In addition, this uniqueness value is larger than the one in \cite{appuni15}, which was about $75\%$ for 2 apps. The difference, as we have discussed in Section \ref{sec:intro}, is due to the fact that only existing apps of users are selected to measure their uniqueness in \cite{appuni15}, meanwhile we optimize the queries for the targeted user's profile and they can also include the queries of non-existing apps.

Finally, we note that similar uniqueness is expected in even larger datasets based on the same reasoning as in \cite{appuni15}. With a growing number of users, the number of available applications should also increase with a lower pace. This means that the number of possible combinations of apps also increases exponentially with each new application added to the dataset.

\subsubsection{TBB users by font detection}
We also run Algorithm \ref{alg:deanon} on the font dataset, described in Section \ref{sec:font_dataset}, to measure the effectiveness of fingerprinting using fonts. We believe that, despite that non-Firefox browsers are not excluded from our dataset, our results are generic. While there are minor differences of JavaScript font detection in different browsers, this method is reliable enough to detect the majority of fonts in all browsers. As a result, a record characterizes the OS of a user rather than his browser. We also studied the uniqueness of fingerprints when they are extended with the screen resolution and available screen size. While the Tor Browser makes various efforts to hinder access to this information, it cannot block full access \cite{tord15}; for example, it warns users of related privacy hazards in case the window of a Tor Browser is maximized.

The distribution of fingerprint length is shown in Fig. \ref{fig:font_lengths}. Results are better for user privacy compared to our app dataset, as only $53.8\%$ of users have fingerprints no longer than the font load limit of $s=10$ imposed by the Tor Browser (see Section \ref{sec:font_app}). Peaks appearing for longer length are likely to be due to pre-set OS configurations that were not altered at all, or only slightly. For example, these can be computers running in institutions or at companies, where configurations are very similar to each other. Barely modified fresh installs could also be behind this phenomena.

\begin{figure}[h!]
    \centering
    \begin{subfigure}[b]{0.3\textwidth}
        \includegraphics[width=\textwidth]{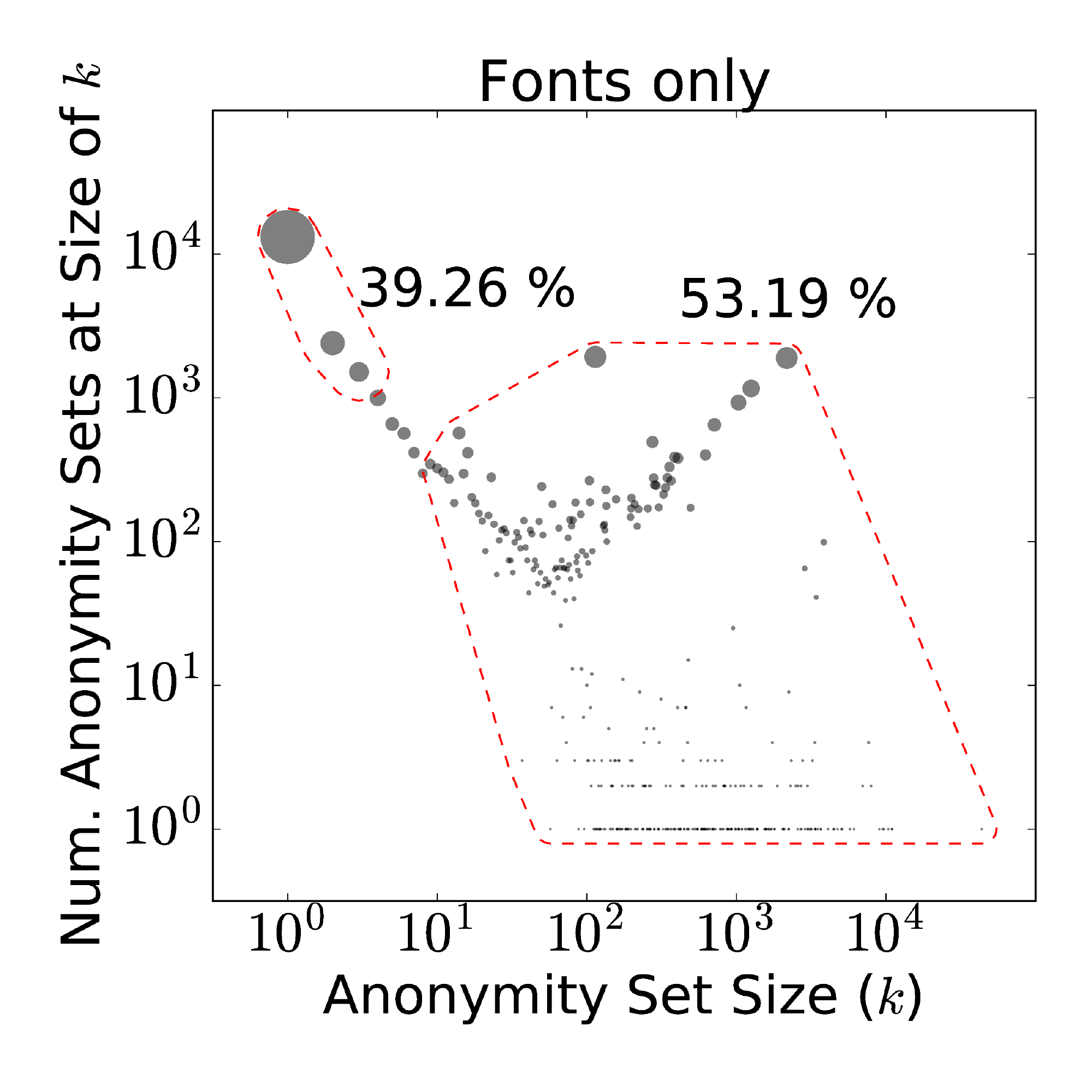}
        \caption{~}
        \label{fig:font_contour:fonts}
    \end{subfigure}
    ~
    \begin{subfigure}[b]{0.3\textwidth}
        \includegraphics[width=\textwidth]{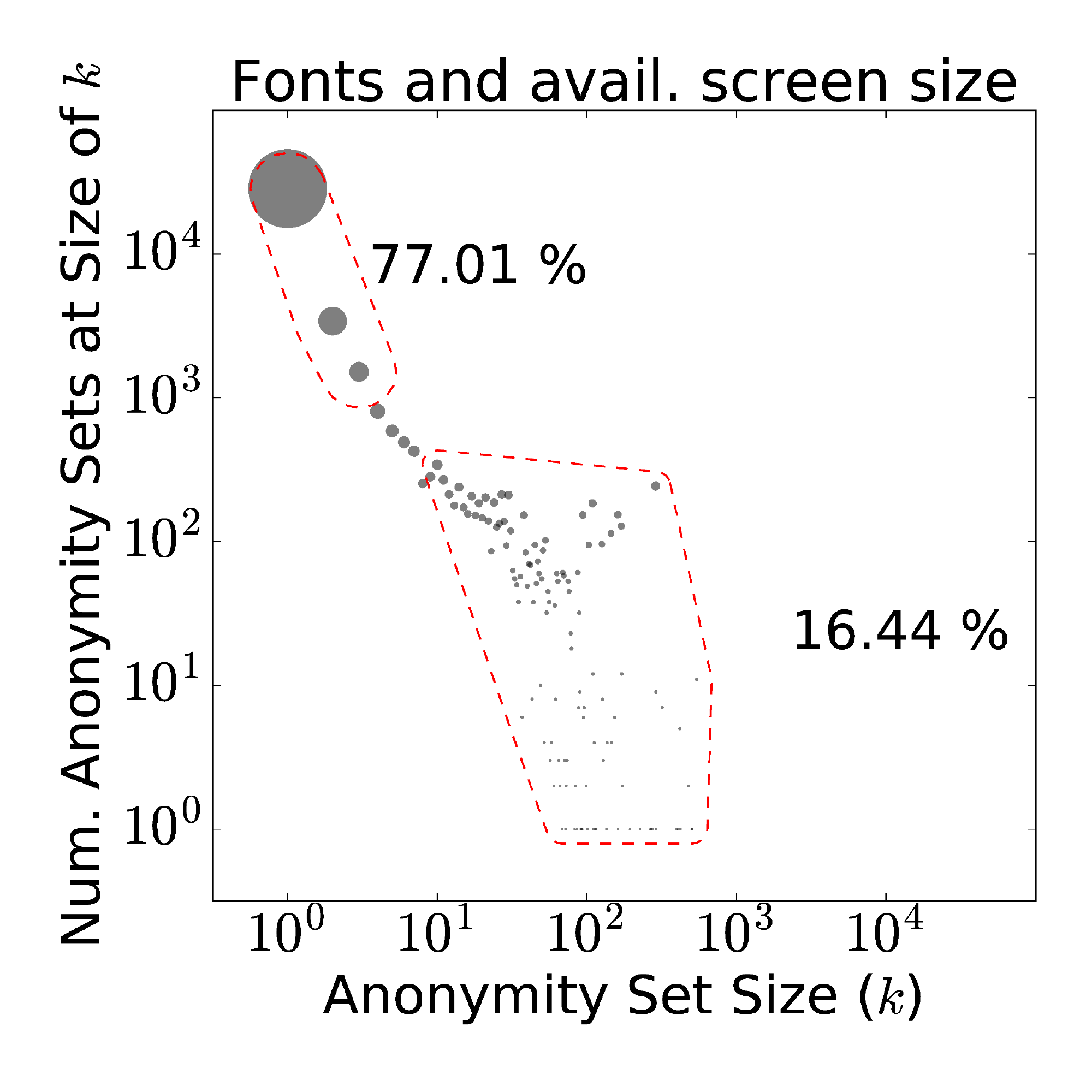}
        \caption{~}
        \label{fig:font_contour:avail}
    \end{subfigure}
    \caption{Comparison of different targeted fingerprinting methods for the font dataset when $s=10$. Even if only fonts are detected (a), a large proportion of users are almost unique (i.e., have anonymity set size at most three). However, adding screen resolution (b) or available screen size (c) greatly increases uniqueness and vanishes large anonymity sets that were visible before.}
    \label{fig:font_contour}
\end{figure}

As font fingerprints are not necessarily unique (due to the same reasons discussed in Section \ref{sec:ind_app}), we report the sizes of anonymity sets for each user. Results are shown in Fig. \ref{fig:font_contour:fonts}. Detection of 10 fonts allows unique identification of $13,155$ users in our datasets ($30.13\%$), and $39.26\%$ of the users are almost unique (i.e., has anonymity set size at most 3). This roughly means that one third of TBB users at a given service can effectively be de-anonymized. 

We also report the size of anonymity sets if screen resolution and available screen size are also added to the fingerprint (see Fig. \ref{fig:font_contour:fonts} and \ref{fig:font_contour:avail}). These results show that these extra attribute values significantly increase the uniqueness of users. While we find that the proportion of almost unique users is roughly doubled when available screen size is added to the fingerprint, probably this would not be the case for TBB users. In particular, TBB users are discouraged to install any browser extensions and other tools which usually change the available screen size.

\begin{figure}[h!]
    \centering
    \begin{subfigure}[b]{0.3\textwidth}
        \includegraphics[width=\textwidth]{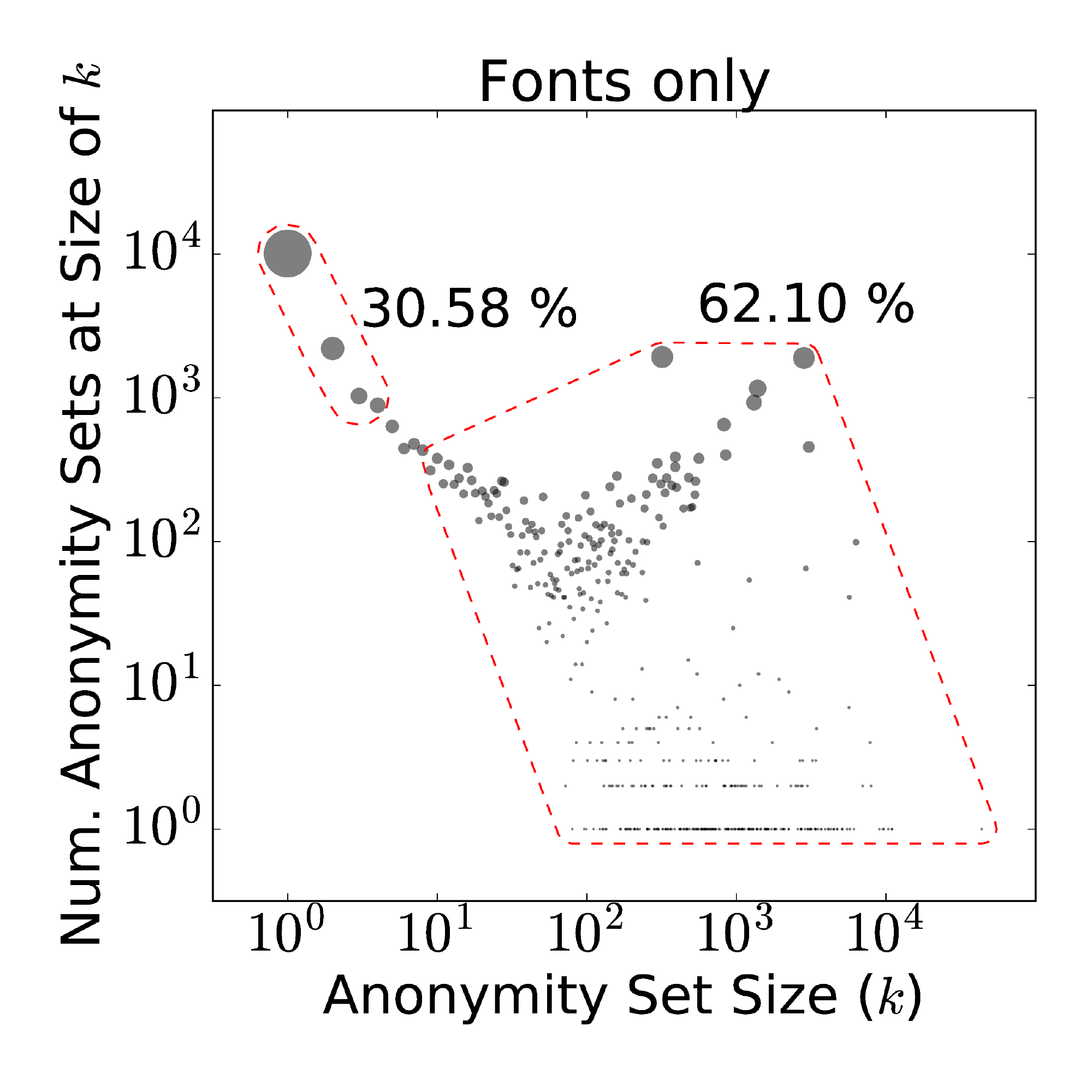}
        \caption{~}
        \label{fig:font_contour2:fonts}
    \end{subfigure}
   ~
    \begin{subfigure}[b]{0.3\textwidth}
        \includegraphics[width=\textwidth]{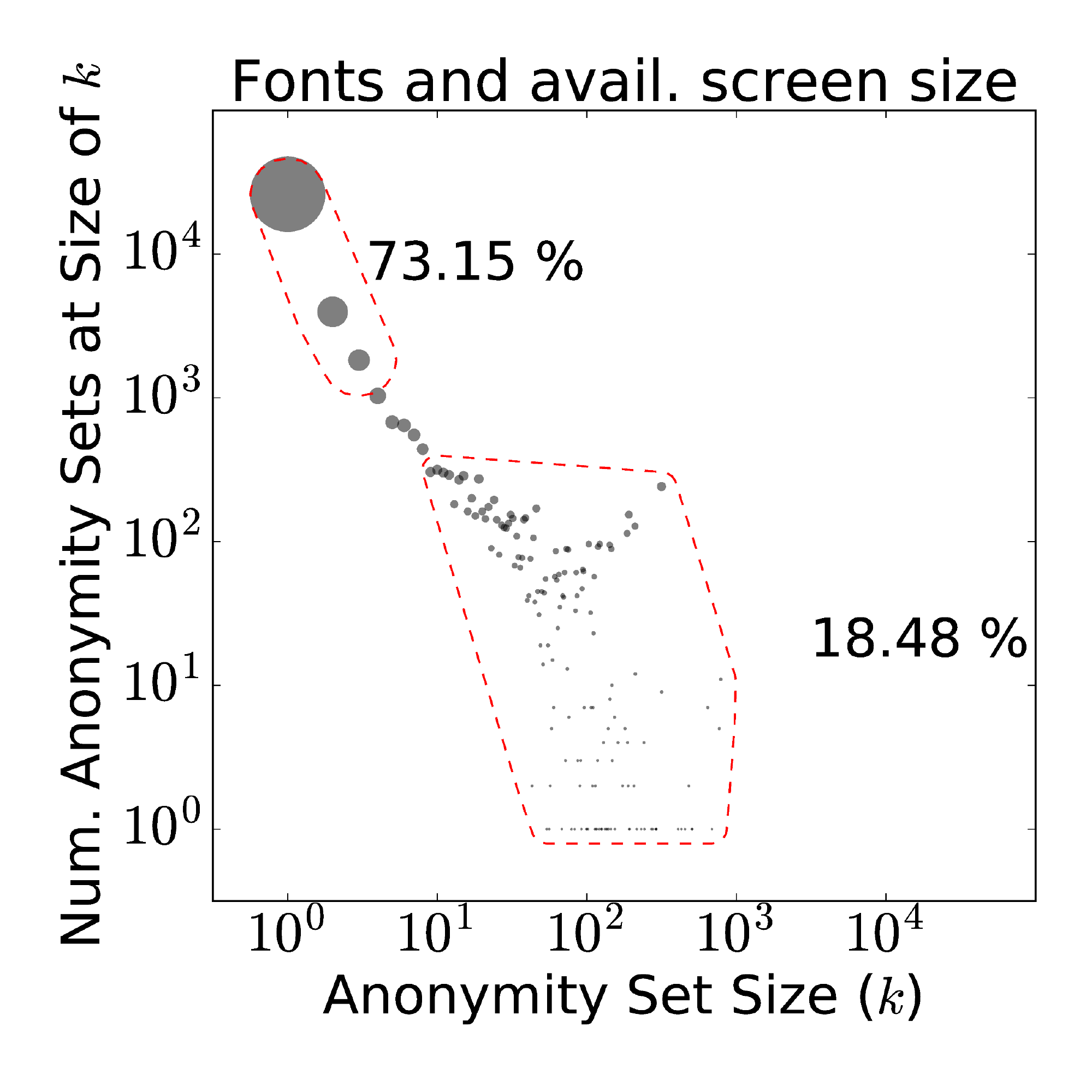}
        \caption{~}
        \label{fig:font_contour2:avail}
    \end{subfigure}
    \caption{Targeted fingerprinting: uniqueness in the font dataset with the size of anonymity sets when $s=5$.}
    \label{fig:font_contour2}
\end{figure}

We additionally examine if decreasing the limit $s$ is a remedy for the situation. Our negative findings show that when the font load limit is decreased to $5$, significant uniqueness can be still observed (see Fig. \ref{fig:font_contour2} for details). For example, detecting solely $5$ fonts makes still $30.58\%$ of users unique (Fig. \ref{fig:font_contour2:fonts}).

\subsection{General fingerprinting}
\begin{figure}[h!]
\begin{minipage}{.4\textwidth}
	\centering
	\includegraphics[width=\textwidth]{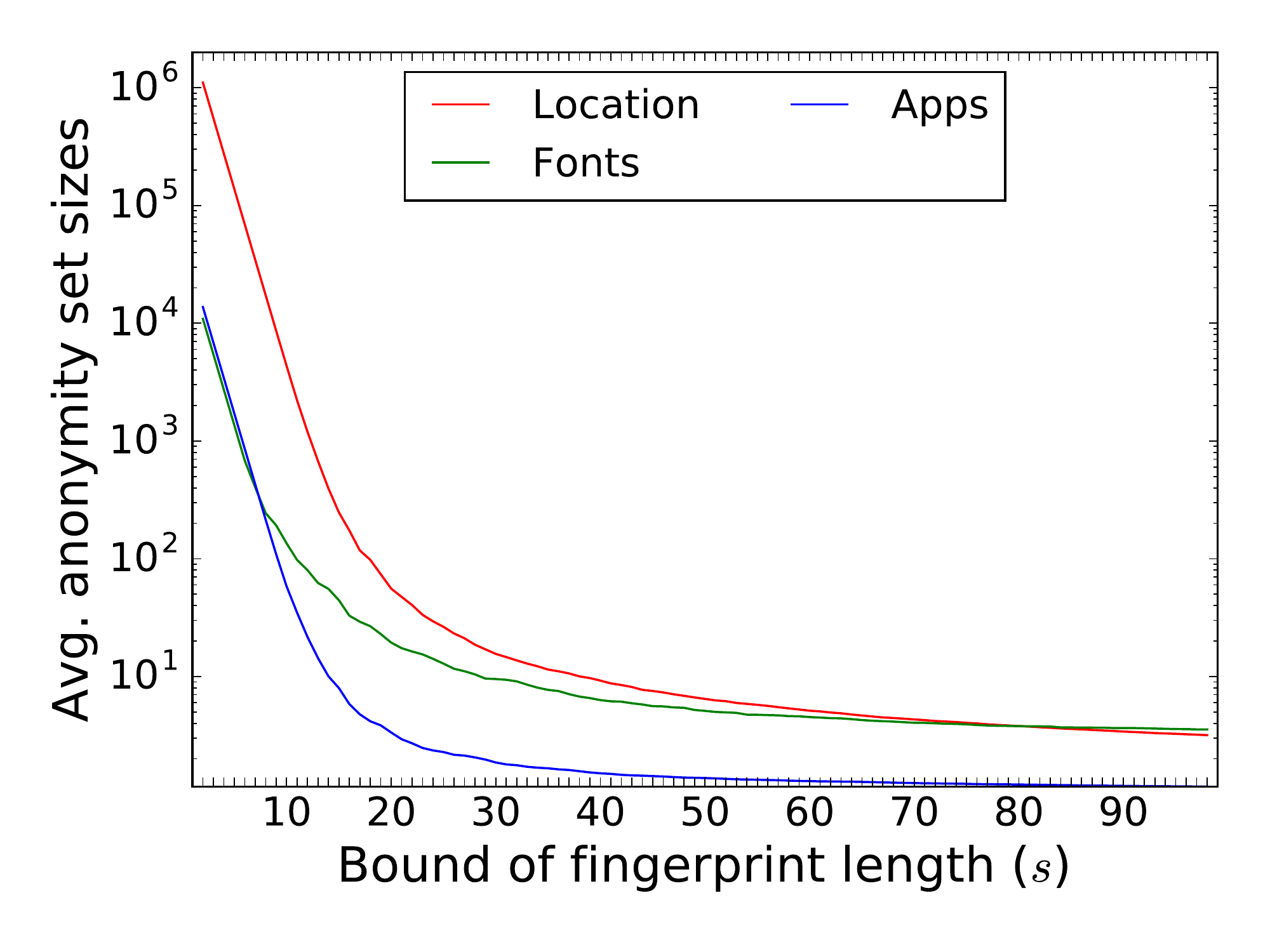}
	\caption{General fingerprinting: average size of anonymity sets as the function of $s$.}
	\label{fig:group_set_size}
\end{minipage}
\begin{minipage}{.4\textwidth}
	\centering
	\includegraphics[width=\textwidth]{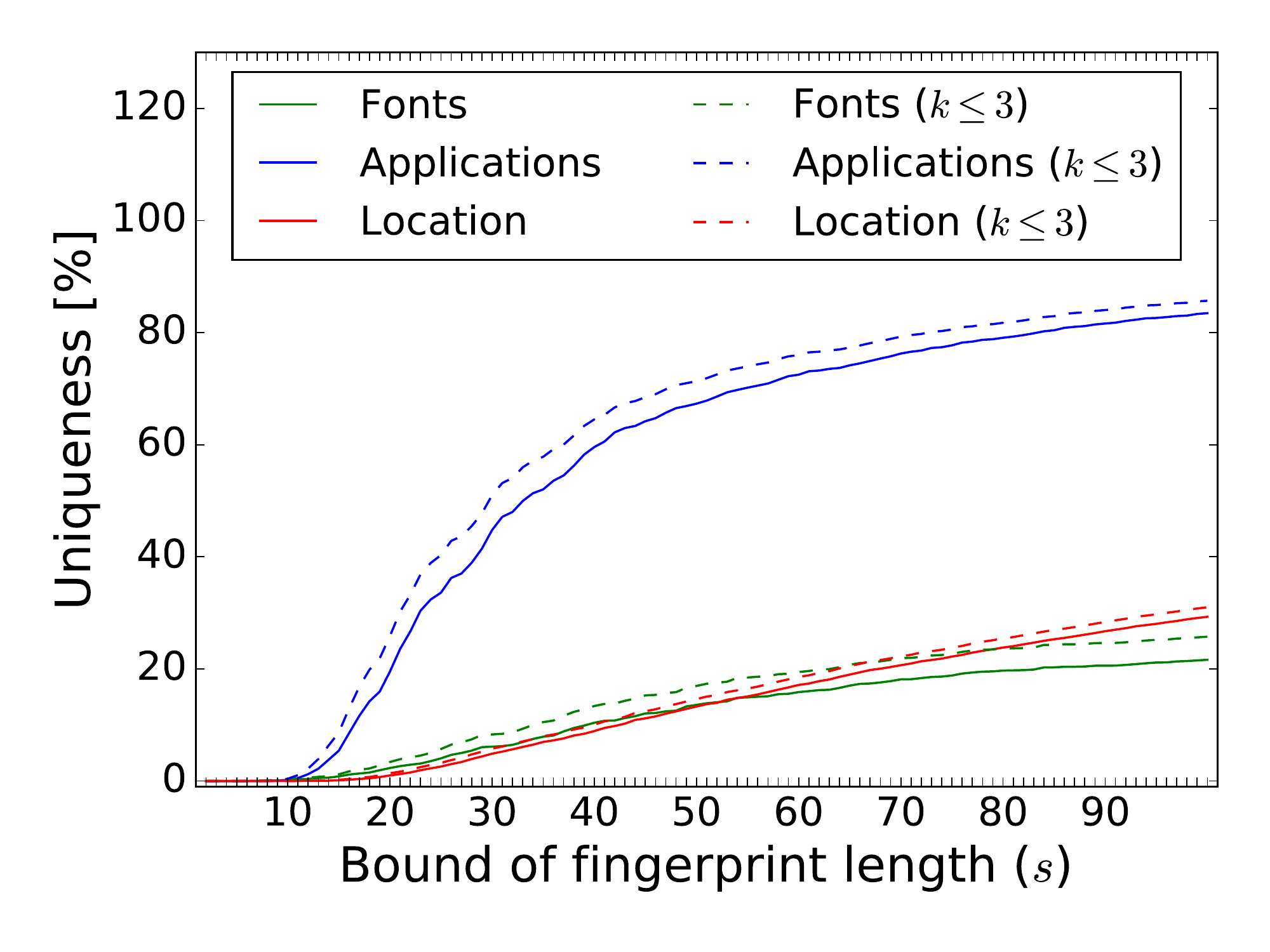}
	\caption{General fingerprinting: proportion of unique and almost unique ($k \leq 3$) users as the function of $s$.}
	\label{fig:uniqueness}
\end{minipage}
\end{figure}

We run Algorithm \ref{alg:link} on the three datasets to compute general fingerprints and partition users into anonymity sets, where users have the same fingerprint value in each set. The average size of anonymity sets for small fingerprint sizes (less than 20) are shown in Fig. \ref{fig:group_set_size}. The average set size drops fast; we observed $135.15$ for $s=10$ in case of the font dataset, $58.21$ and $4344.93$  for the app and location dataset, respectively. In case of the application dataset, the average anonymity set size is $1.96$ at $s=29$. In case of the other datasets, the average never falls below $3$, and we found $3.99$ at $s=76$ for the app dataset, and $3.96$ at $s=73$ for the font dataset.

While these results may not be promising to the adversary, measuring the proportion of unique users, whose anonymity sets are singletons, provides a different view. Unique users do not share the same fingerprint value with any other users in the dataset.
As Fig. \ref{fig:uniqueness} shows, the proportion of unique users remains relatively low. The proportion does not change too much even for almost unique users which have anonymity set with size at most 3.

\begin{figure*}[h!]
    \centering
    \begin{subfigure}[b]{0.30\textwidth}
        \includegraphics[width=\textwidth]{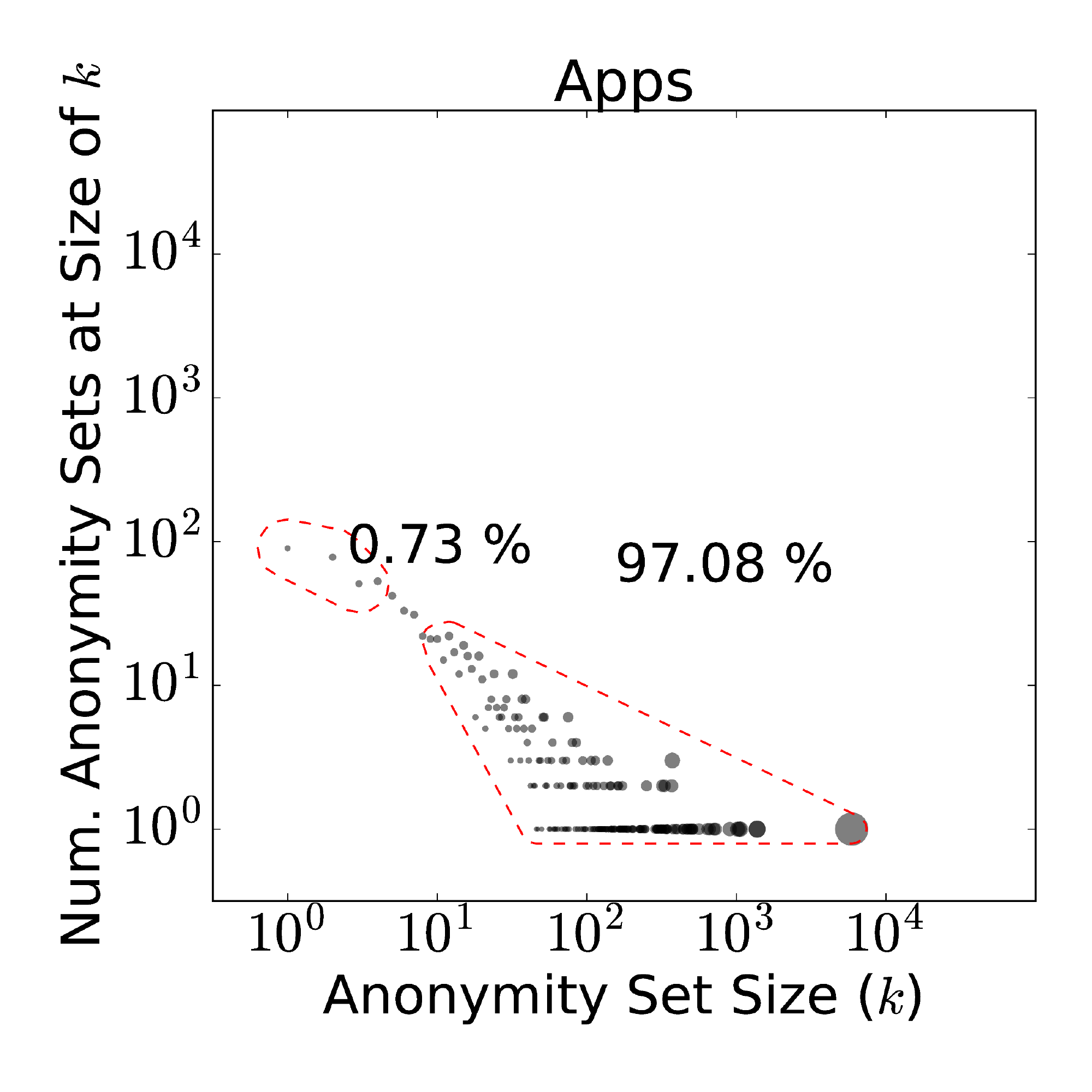}
        \caption{$s=10$}
        \label{fig:apps_diffk:10}
    \end{subfigure}
    ~
    \begin{subfigure}[b]{0.30\textwidth}
        \includegraphics[width=\textwidth]{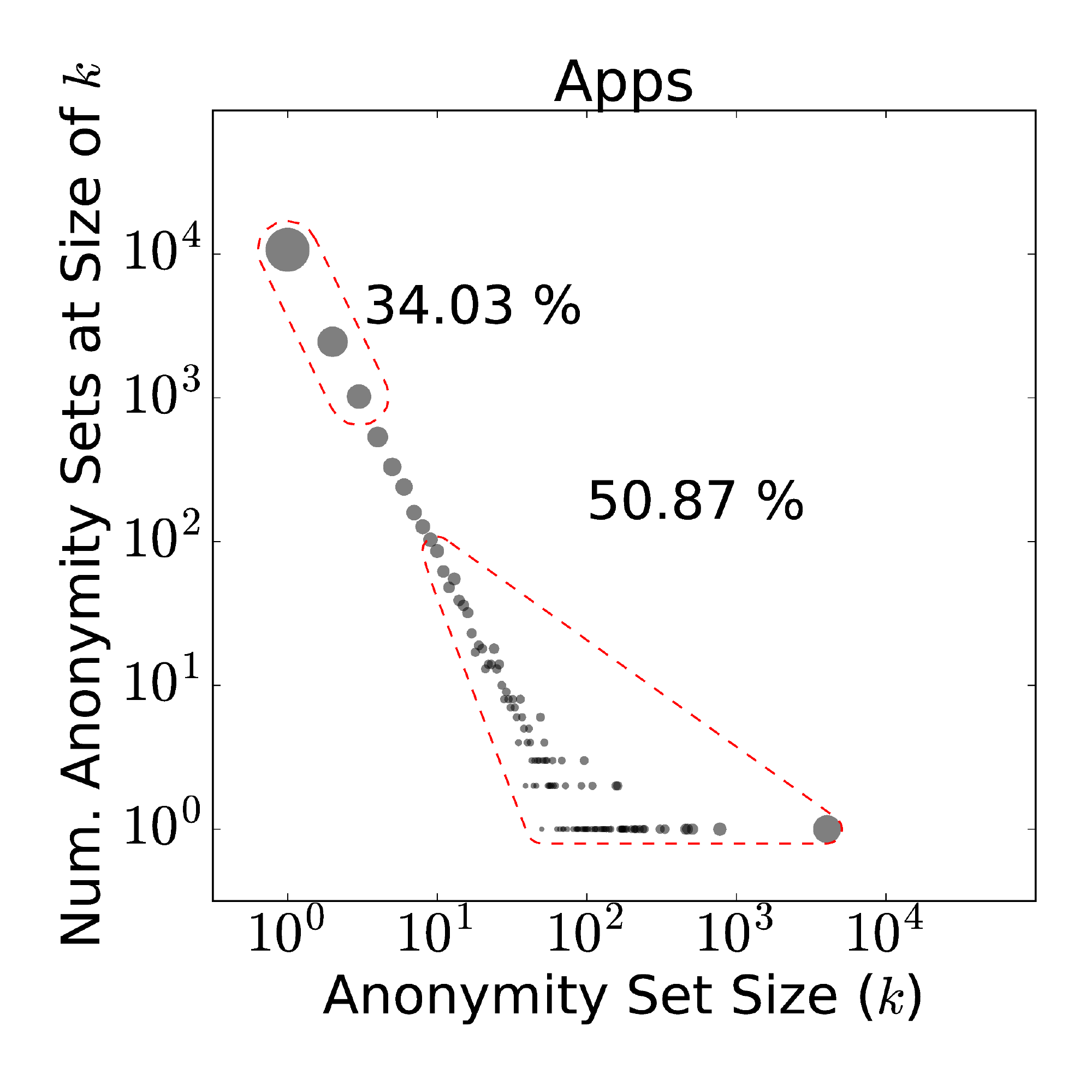}
        \caption{$s=20$}
        \label{fig:apps_diffk:20}
    \end{subfigure}
    ~
    \begin{subfigure}[b]{0.30\textwidth}
        \includegraphics[width=\textwidth]{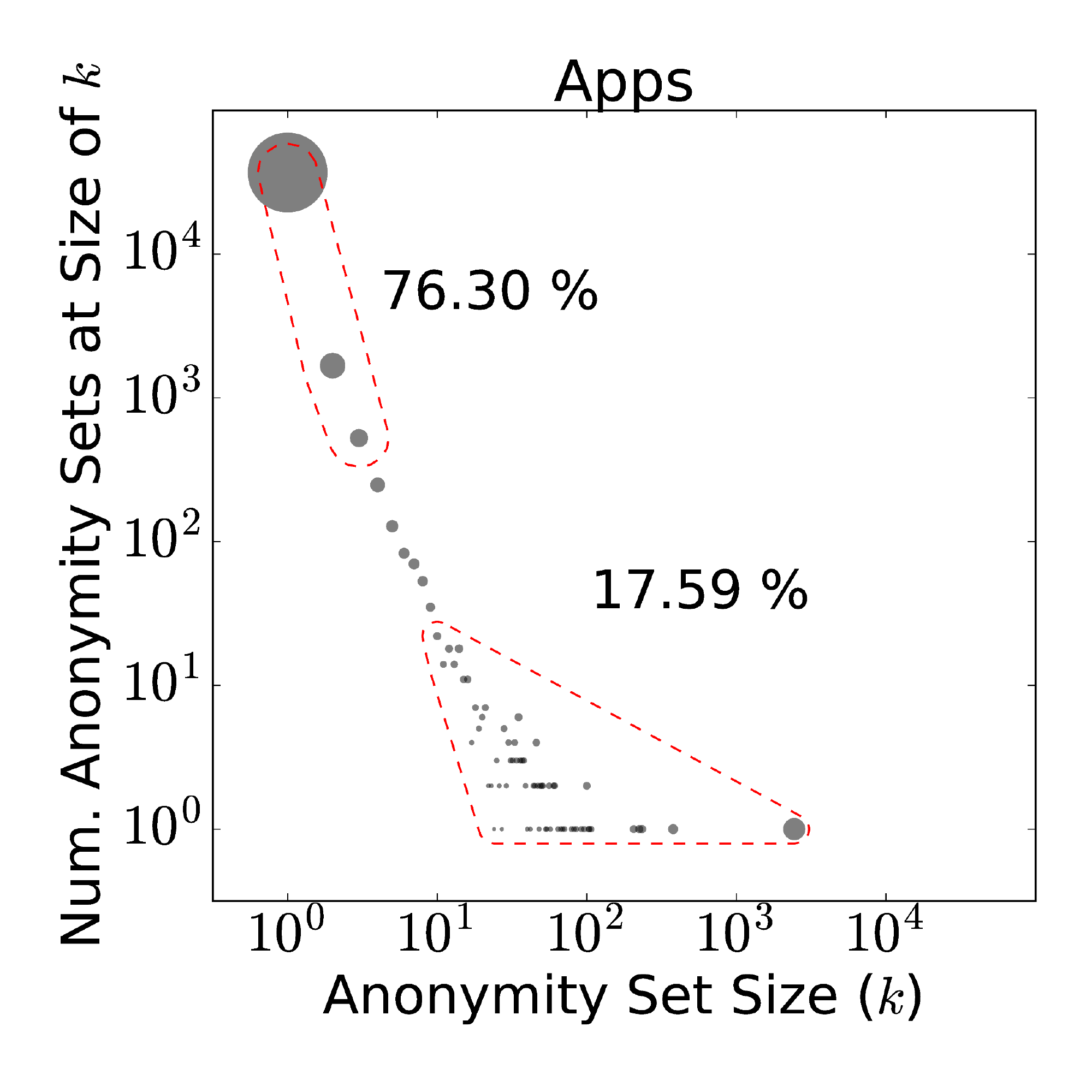}
        \caption{$s=50$}
        \label{fig:apps_diffk:50}
    \end{subfigure}
    \caption{General fingerprinting: distribution of size of anonymity sets for different values of $s$ on the apps dataset.}
    \label{fig:apps_diffk}
\end{figure*}

\begin{figure*}[h!]
    \centering
    \begin{subfigure}[b]{0.30\textwidth}
        \includegraphics[width=\textwidth]{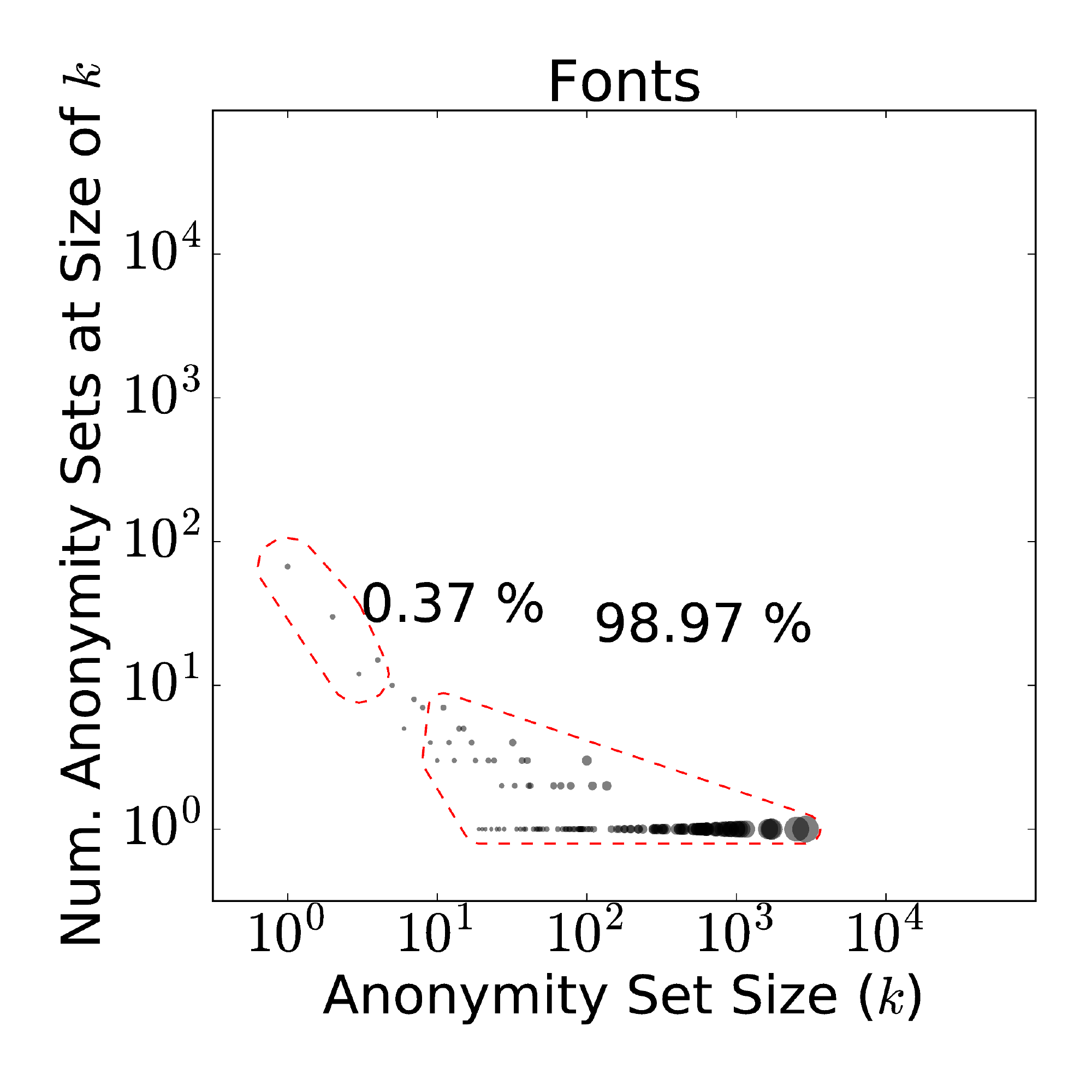}
        \caption{$s=10$}
        \label{fig:font_diffk:10}
    \end{subfigure}
    ~
    \begin{subfigure}[b]{0.30\textwidth}
        \includegraphics[width=\textwidth]{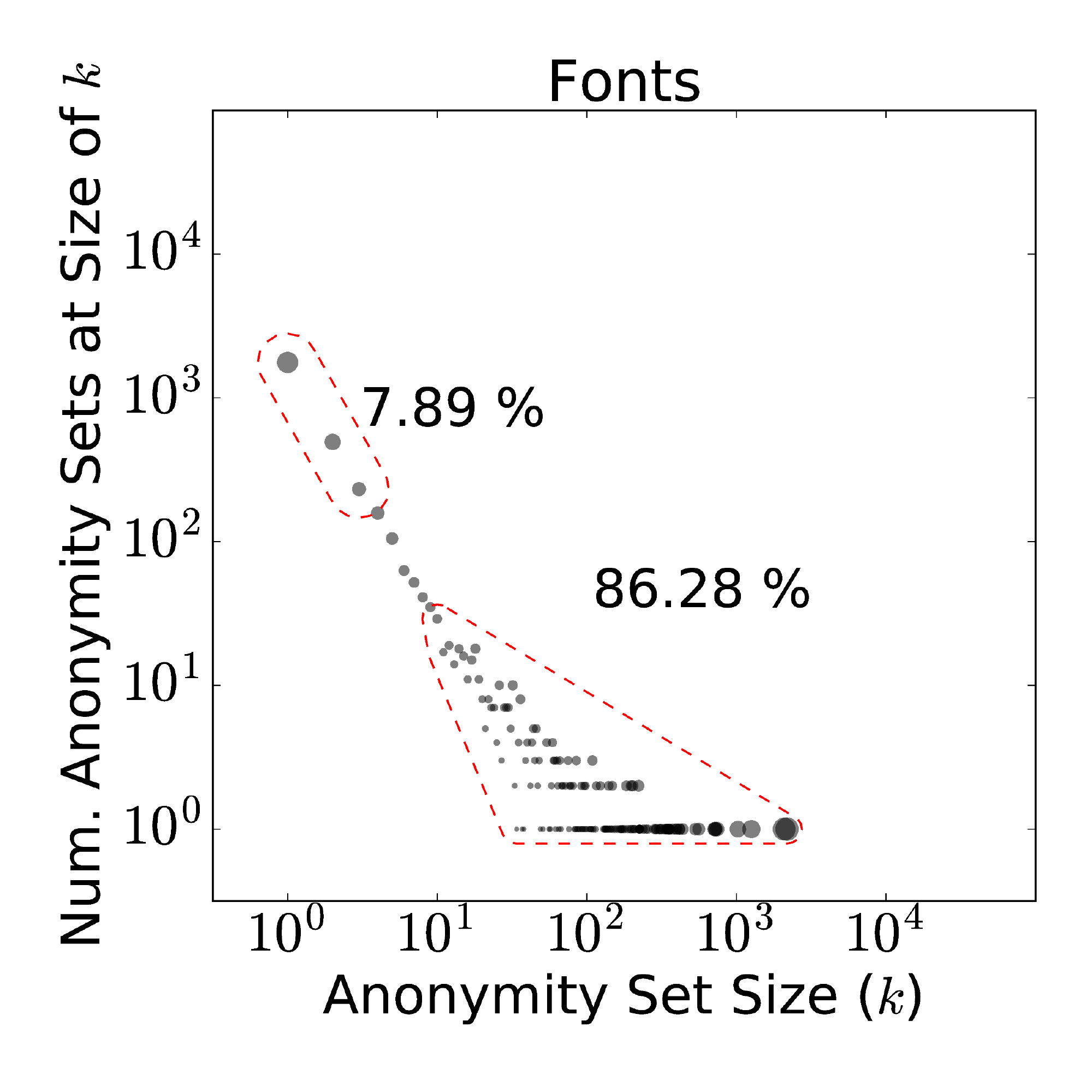}
        \caption{$s=25$}
        \label{fig:font_diffk:25}
    \end{subfigure}
    ~
    \begin{subfigure}[b]{0.30\textwidth}
        \includegraphics[width=\textwidth]{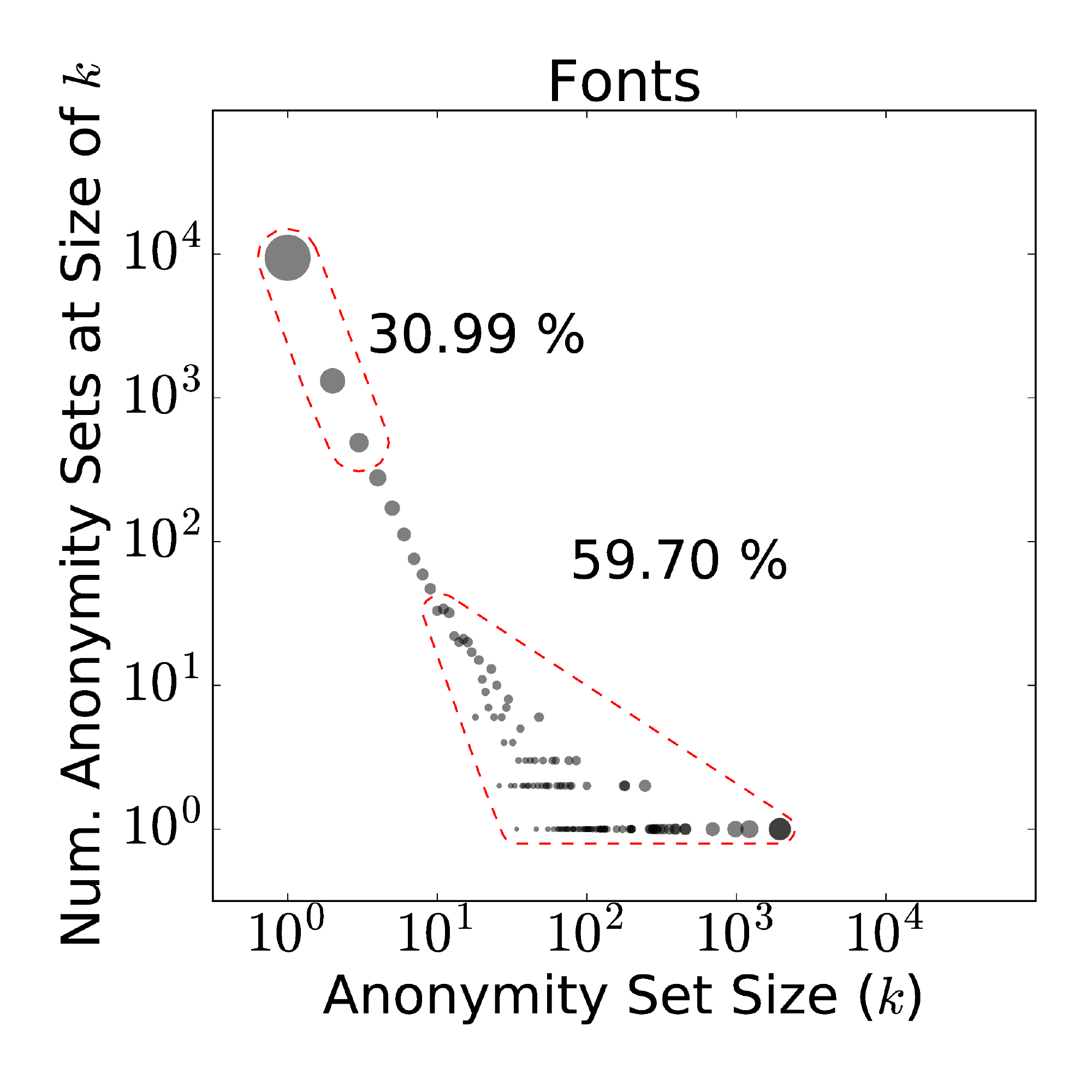}
        \caption{$s=100$}
        \label{fig:font_diffk:100}
    \end{subfigure}
    \caption{General fingerprinting: distribution of size of anonymity sets for different values of $s$ on the font dataset.}
    \label{fig:font_diffk}
\end{figure*}

\begin{figure*}[h!]
    \centering
    \begin{subfigure}[b]{0.30\textwidth}
        \includegraphics[width=\textwidth]{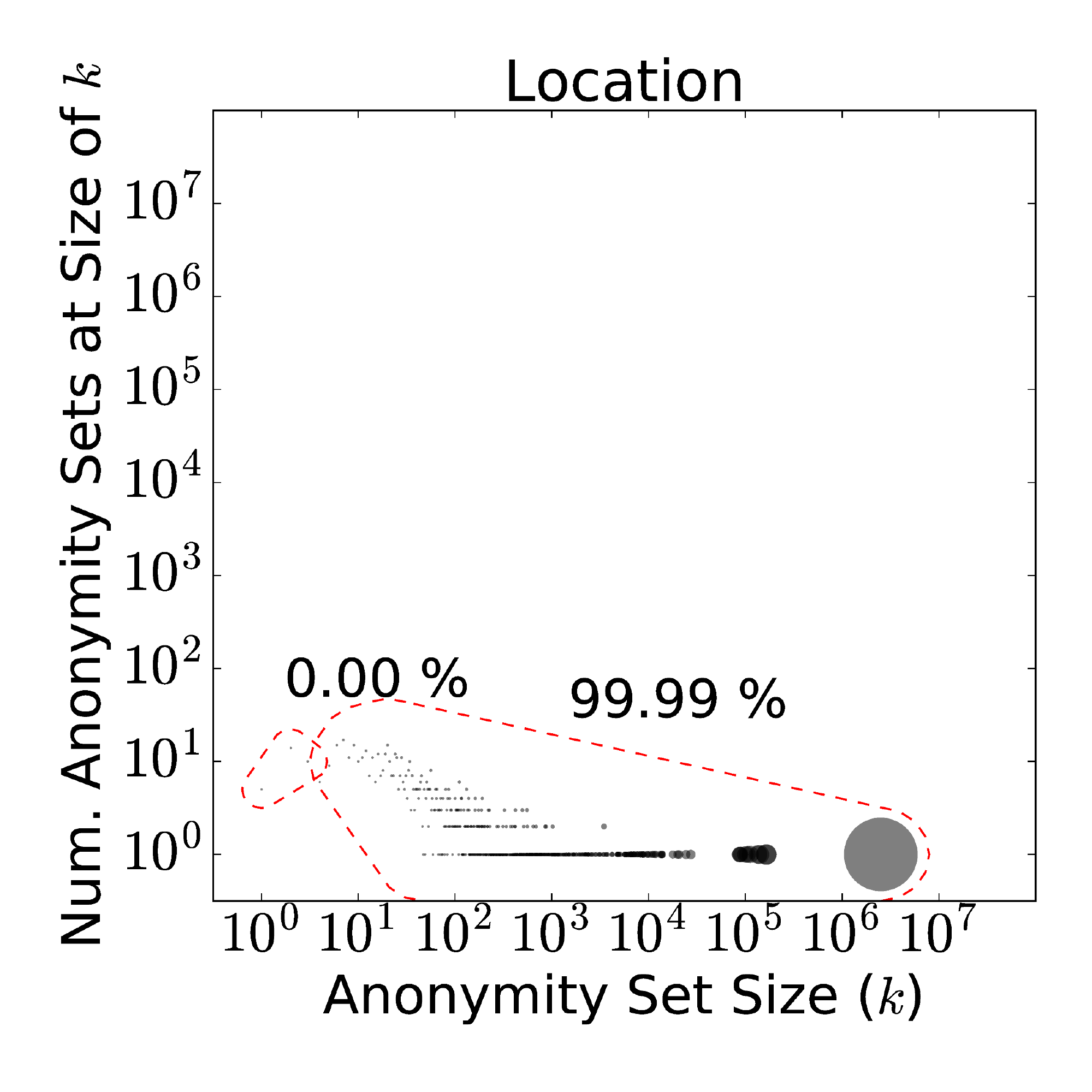}
        \caption{$s=10$}
        \label{fig:loc_diffk:10}
    \end{subfigure}
    ~
    \begin{subfigure}[b]{0.30\textwidth}
        \includegraphics[width=\textwidth]{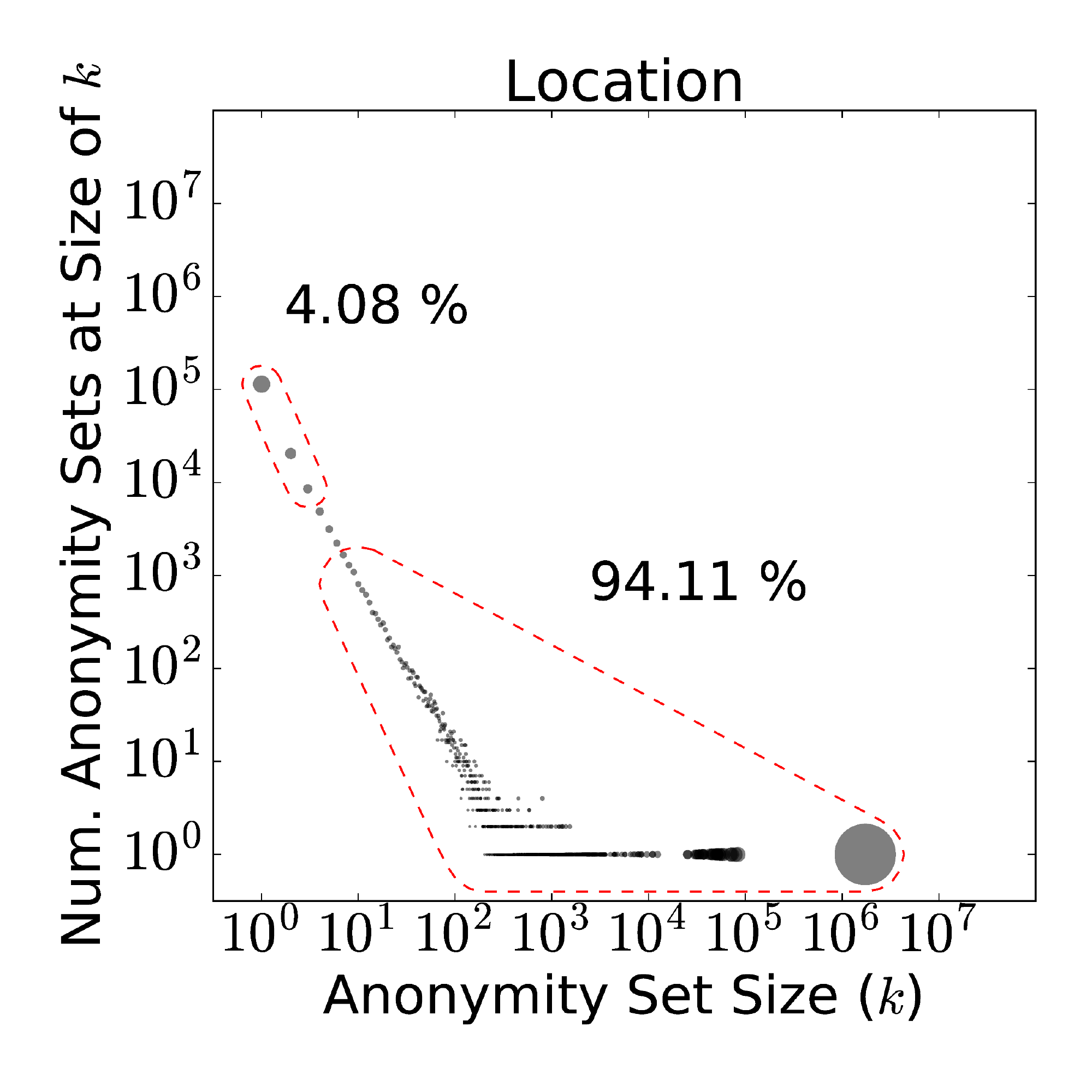}
        \caption{$s=25$}
        \label{fig:loc_diffk:25}
    \end{subfigure}
    ~
    \begin{subfigure}[b]{0.30\textwidth}
        \includegraphics[width=\textwidth]{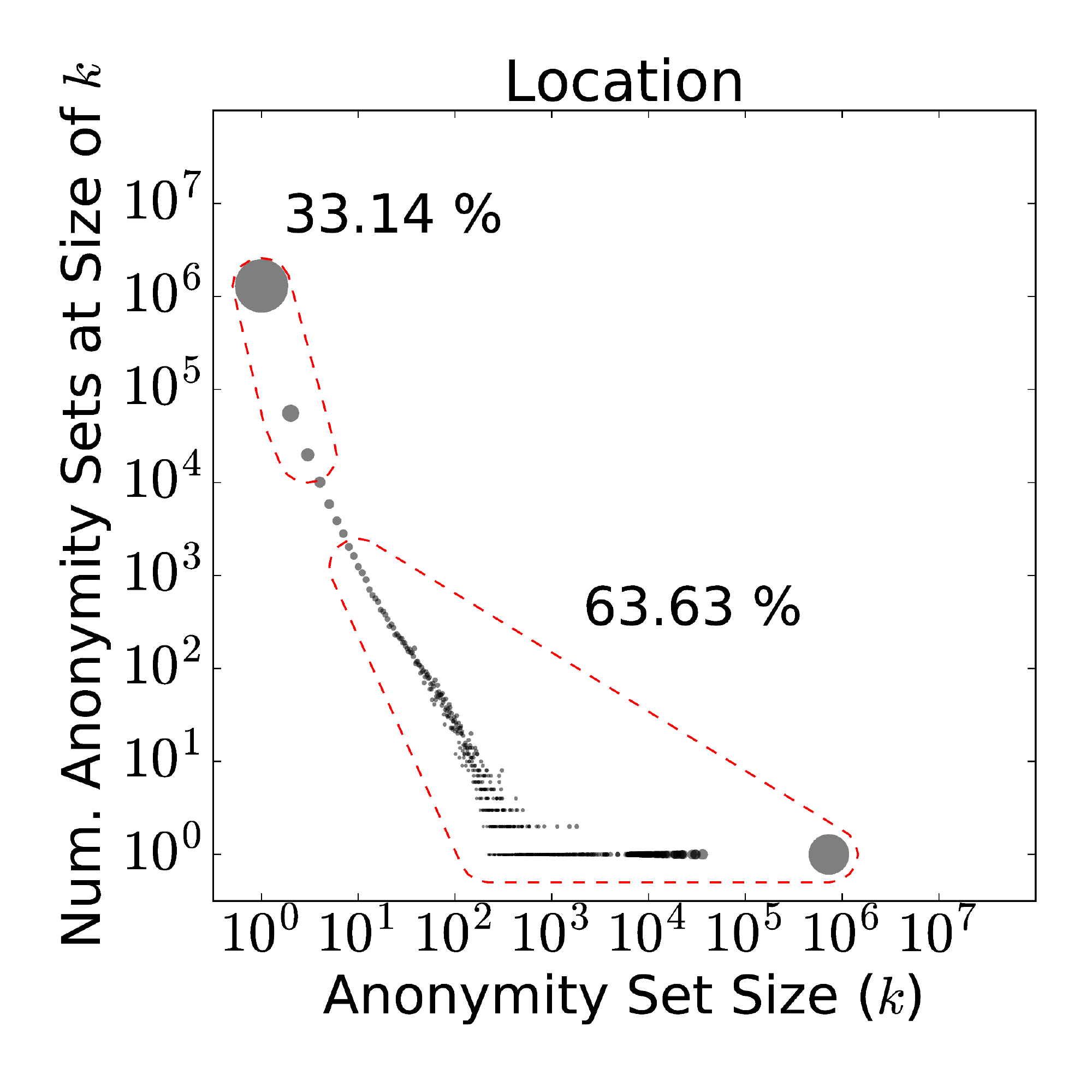}
        \caption{$s=100$}
        \label{fig:loc_diffk:100}
    \end{subfigure}
    \caption{General fingerprinting: distribution of size of anonymity sets for different values of $s$ on the location dataset.}
    \label{fig:location_diffk}
\end{figure*}

\begin{figure*}[h!]
    \centering
    \begin{subfigure}[b]{0.30\textwidth}
        \includegraphics[width=\textwidth]{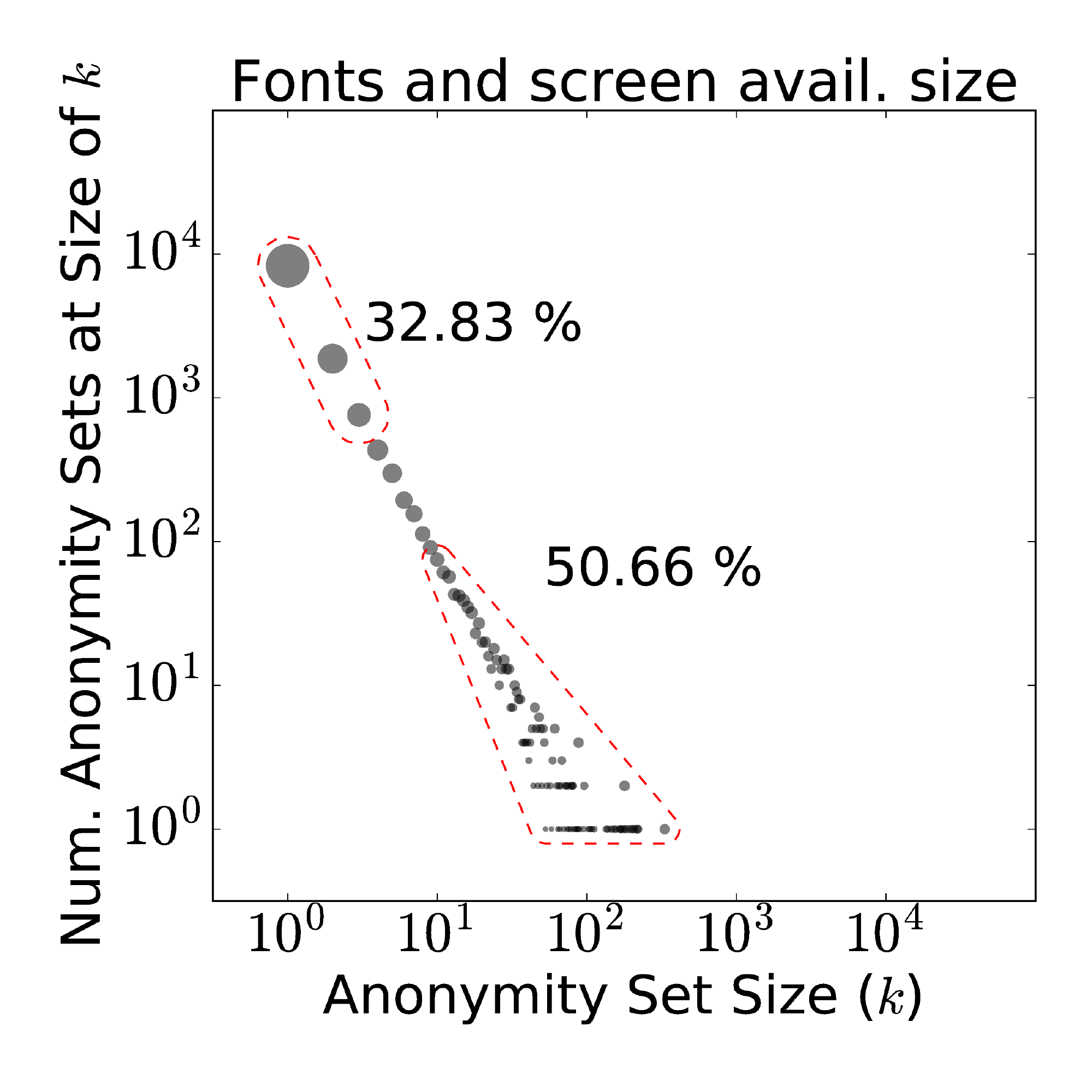}
        \caption{$s=10$}
        \label{fig:loc_diffk:10}
    \end{subfigure}
    ~
    \begin{subfigure}[b]{0.30\textwidth}
        \includegraphics[width=\textwidth]{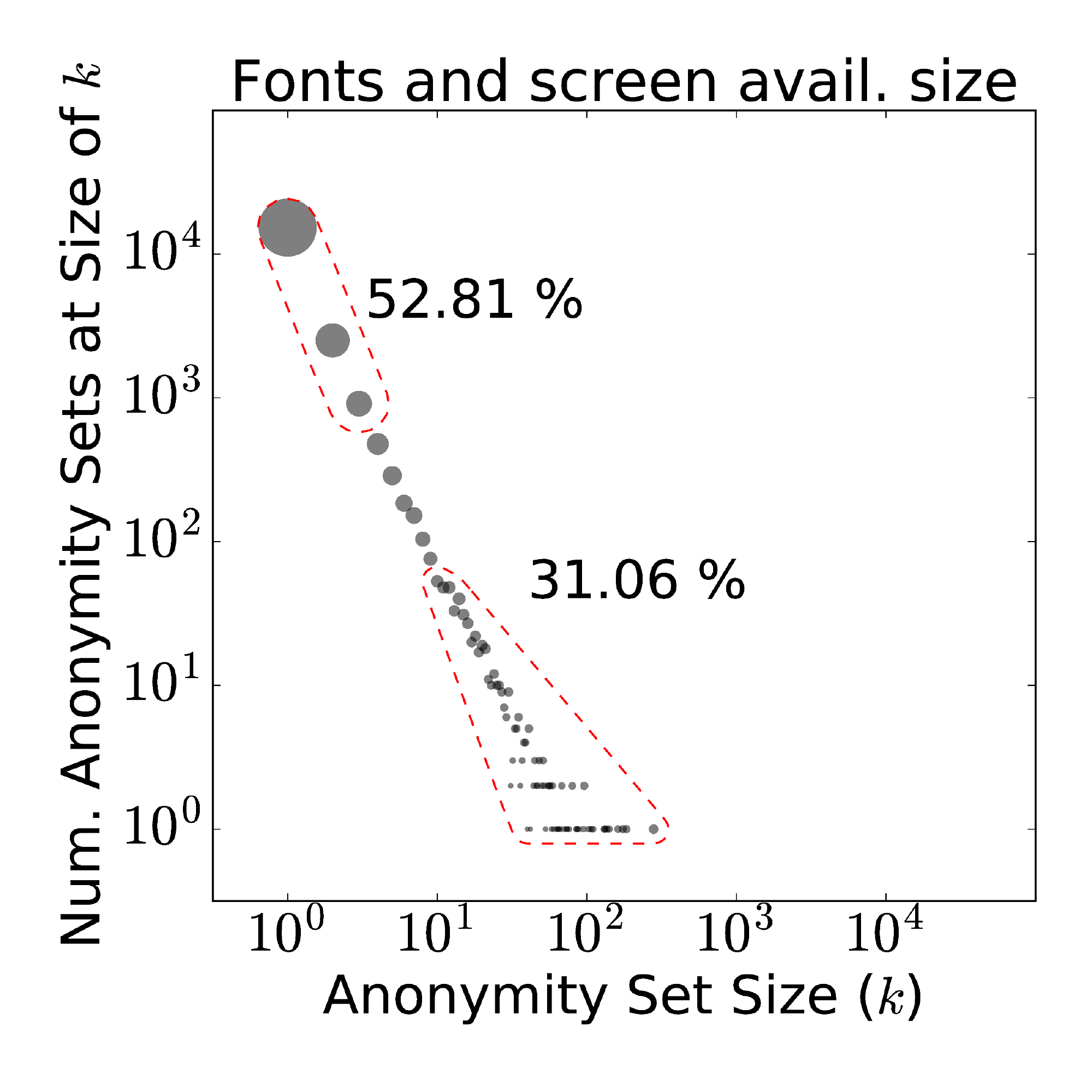}
        \caption{$s=25$}
        \label{fig:loc_diffk:25}
    \end{subfigure}
    ~
    \begin{subfigure}[b]{0.30\textwidth}
        \includegraphics[width=\textwidth]{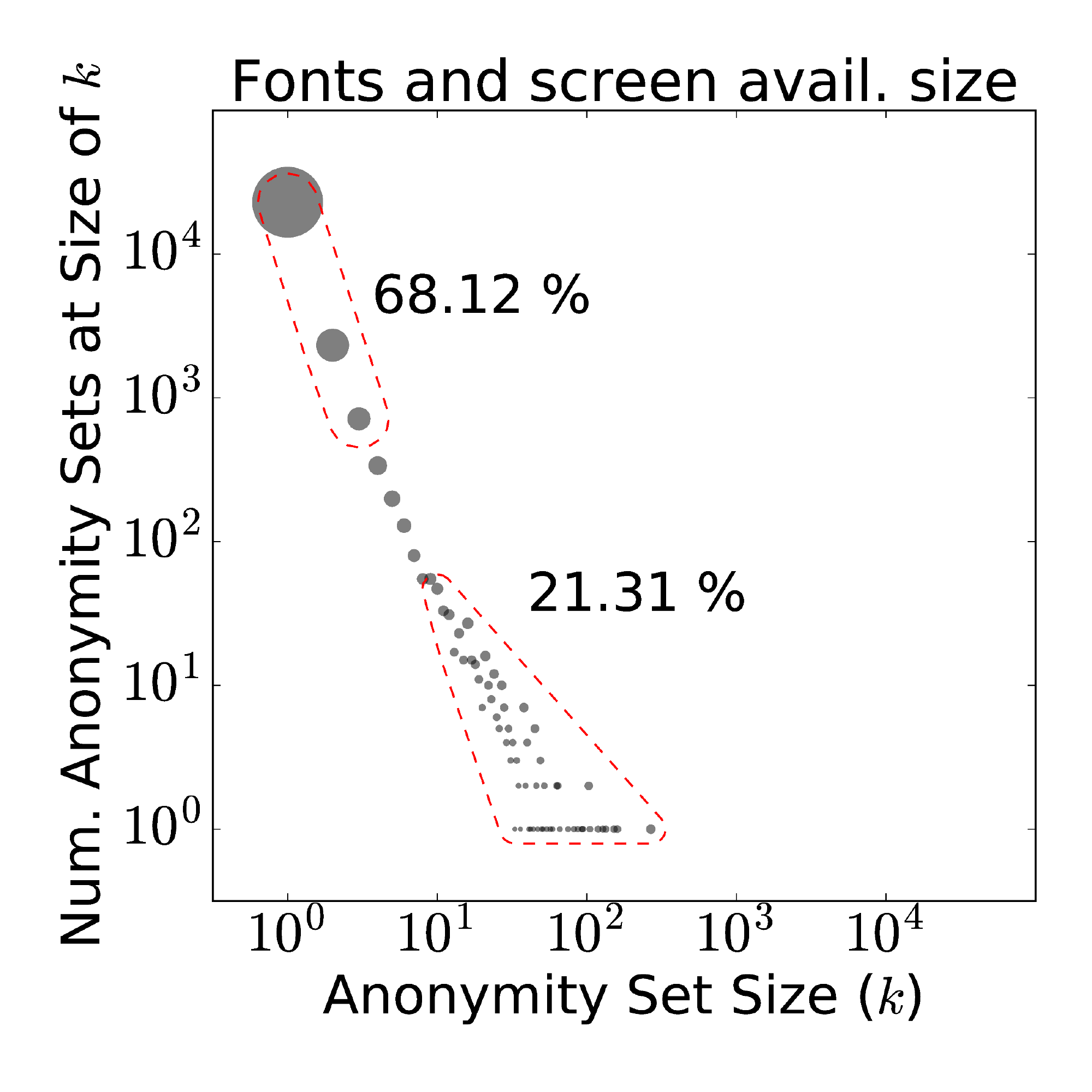}
        \caption{$s=100$}
        \label{fig:loc_diffk:100}
    \end{subfigure}
    \caption{General fingerprinting: distribution of size of anonymity sets for different values of $s$ on the font dataset (with screen resolution and available screen size).}
    \label{fig:font_avail_diffk}
\end{figure*}

As regards linking attack against TBB users, we find that the available screen size as an auxiliary information (in addition to the detection of 10 fonts) makes this attack a significant threat, as $8,303$ users ($19.02\%$) are unique in this case. 
Without available screen size, less than 0.37\% of users are unique with $s=10$, but 7.89\% of users are almost unique with $s=25$ (Fig. \ref{fig:font_diffk}; with available screen size in Fig. \ref{fig:font_avail_diffk}). 
We provide further details in Fig. \ref{fig:apps_diffk} on the apps dataset, where uniqueness was the highest, that is, 76.30\% of users have a size of anonymity set smaller than 4. This means that even if all 50 apps need to be pre-defined for all users, which is the case for iOS 9 at present, privacy threats still persist. For the location dataset, we find that 29\% of all users are unique, and the average size of the anonymity set is 3.13 with $s=100$ (Fig. \ref{fig:location_diffk}). That is, an authority would need to deploy at least 100 cameras in the city to re-identify every third person on average in an pseudonymized CDR dataset (see Section \ref{sec:app_location}).

\section{Conclusion}
In this paper we addressed two fingerprinting problems with constraints on the maximum size of the fingerprint. Targeted fingerprinting is executed on a specific user, whose profile and identity are known to the adversary, and the goal is to re-identify him/her in an anonymous context, where only a limited number of his attributes can be queried. General fingerprinting is used to link the activities of multiple users without their re-identification in an anonymous context, where only a limited number of their attributes can be queried. We provided essentially the best fingerprinting algorithms for both problems.

We evaluated these fingerprinting methods on real-life examples focusing on web tracking, smartphone user identification and mass de-anonymization of location datasets. We showed that targeted fingerprinting is a significant threat to user privacy, as a significant ratio of users can be re-identified with our technique in case of web tracking and smartphone identification. This holds even when tighter constraints are applied than the default. Our results with general fingerprinting also support this claim; we found that $19.02\%$ of web browsers could be uniquely identified by combining the presence of 10 fonts and the available screen size. We also found that by installing 100 surveillance cameras in a city of 4 million people, one can re-identify on average every fourth person in pseudonymized  CDR dataset. Moreover, by querying 50 apps on a smartphone, which can be even pre-defined and identical for all users, one can obtain a unique fingerprint value for more than 65\% of users in a dataset of 55k users.

All these findings show a conceptual weakness of limiting attribute access as a privacy guard.

\section{Acknowledgement}
The authors are thankful for Viktor F\'{a}bi\'{a}n and Jagdish Prasad Achara for the discussions and their advises related to the iOS platform. This work was carried out during the tenure of an ERCIM ’Alain Bensoussan‘ Fellowship Programme. Some icons we used in the illustrations were made by Freepik (from \url{www.flaticon.com})

\bibliographystyle{abbrv}

\small
\bibliography{references}

\begin{thebibliography}{10}

\bibitem{GDPR}
{European Commission. Proposal for European Parliament and the Council (General
  Data Protection Regulation)}, 2012.
\newblock
  \url{http://eur-lex.europa.eu/LexUriServ/LexUriServ.do?uri=COM:2012:0011:FIN:EN:PDF}.

\bibitem{appleios9}
Apple uikit framework reference, entry on canopenurl call.
\newblock
  \url{https://developer.apple.com/library/ios/documentation/UIKit/Reference/UIApplication_Class/#//apple_ref/occ/instm/UIApplication/canOpenURL:},
  2016.

\bibitem{Acar:2014}
G.~Acar, C.~Eubank, S.~Englehardt, M.~Juarez, A.~Narayanan, and C.~Diaz.
\newblock The web never forgets: Persistent tracking mechanisms in the wild.
\newblock In {\em ACM CCS}, pages 674--689, 2014.

\bibitem{appuni15}
J.~P. Achara, G.~Acs, and C.~Castelluccia.
\newblock On the unicity of smartphone applications.
\newblock In {\em Proceedings of WPES}, pages 27--36. ACM, 2015.

\bibitem{Art29op}
{Article 29 Data Protection Working Party}.
\newblock Opinion 05/2014 on anonymization techniques, April 2014.

\bibitem{cross12}
K.~Boda, A.~M. F\"{o}ldes, G.~G. Guly\'{a}s, and S.~Imre.
\newblock User tracking on the web via cross-browser fingerprinting.
\newblock In P.~Laud, editor, {\em Information Security Technology for
  Applications}, volume 7161 of {\em LNCS}, pages 31--46. 2012.

\bibitem{Bujlow:2015}
T.~Bujlow, V.~Carela-Espanol, J.~Sole-Pareta, and P.~Barlet-Ros.
\newblock Web tracking: Mechanisms, implications, and defenses.
\newblock In {\em \url{http://arxiv.org/abs/1507.07872}}, 2015.

\bibitem{Cai:2012}
X.~Cai, X.~C. Zhang, B.~Joshi, and R.~Johnson.
\newblock Touching from a distance: Website fingerprinting attacks and
  defenses.
\newblock In {\em ACM CCS}, 2012.

\bibitem{Chekuri04maximumcoverage}
C.~Chekuri and A.~Kumar.
\newblock Maximum coverage problem with group budget constraints and
  applications.
\newblock In {\em Approximation, Randomization, and Combinatorial
  Optimization}, pages 72--83. Springer LNCS, 2004.

\bibitem{Nature13}
Y.-A. de~Montjoye, C.~A. Hidalgo, M.~Verleysen, and V.~D. Blondel.
\newblock Unique in the crowd: The privacy bounds of human mobility.
\newblock {\em Scientific Reports, Nature}, March 2013.

\bibitem{Science15}
Y.-A. de~Montjoye, L.~Radaelli, V.~K. Singh, and A.~Pentland.
\newblock Unique in the shopping mall: On the reidentifiability of credit card
  metadata.
\newblock {\em Science}, 347(6221), January 2015.

\bibitem{pan10}
P.~Eckersley.
\newblock How unique is your web browser?
\newblock In {\em PETS}, pages 1--18, 2010.

\bibitem{Feige98}
U.~Feige.
\newblock A threshold of $\ln(n)$ for approximating set cover.
\newblock {\em Journal of the ACM}, 45(4):634--652, July 1998.

\bibitem{torframe13}
gacar.
\newblock Improve persistence and webfont compatibility of font patch, comment
  \#13.
\newblock
  \url{https://trac.torproject.org/projects/tor/ticket/5798comment:\#13}, 2013.

\bibitem{torstats14}
M.~Graham and S.~D. Sabbata.
\newblock The anonymous internet.
\newblock \url{http://geography.oii.ox.ac.uk/?page=tor}, 2014.

\bibitem{tortrack15}
G.~G. Guly\'{a}s, G.~Acs, and C.~Castelluccia.
\newblock Update your tor browser settings - otherwise it is less anonymous
  than you would think.
\newblock
  \url{https://gulyas.info/blog/read/16/2015-11-30-Update-your-TOR-Browser-settings-otherwise-it-is-less-anonymous-than-you-would-think.php},
  11 2015.

\bibitem{Danezis:2015}
J.~Hayes and G.~Danezis.
\newblock k-fingerprinting: a robust scalable website fingerprinting technique.
\newblock In {\em \url{http://arxiv.org/abs/1509.00789}}, 2015.

\bibitem{LemireBK14}
D.~Lemire, L.~Boytsov, and N.~Kurz.
\newblock {SIMD} compression and the intersection of sorted integers.
\newblock {\em CoRR}, abs/1401.6399, 2014.

\bibitem{twitter14}
J.~Marshall.
\newblock Twitter is tracking users’ installed apps for ad targeting.
\newblock
  \url{http://blogs.wsj.com/cmo/2014/11/26/twitter-is-tracking-users-installed-apps-for-ad-targeting/},
  11 2014.

\bibitem{tord15}
E.~C. Mike~Perry and S.~Murdoch.
\newblock The design and implementation of the tor browser [draft].
\newblock \url{https://www.torproject.org/projects/torbrowser/design/}, 5 2015.

\bibitem{masking_vldb}
R.~Motwani and Y.~Xu.
\newblock Efficient algorithms for masking and finding quasi-identifiers.
\newblock In {\em VLDB}, 2007.

\bibitem{Nemhauser78}
G.~Nemhauser, L.~Wolsey, and M.~Fisher.
\newblock An analysis of approximations for maximizing submodular set functions
  i.
\newblock {\em Mathematical Programming}, 14(1):265--294, 1978.

\bibitem{cookieless13}
N.~Nikiforakis, A.~Kapravelos, W.~Joosen, C.~Kruegel, F.~Piessens, and
  G.~Vigna.
\newblock Cookieless monster: Exploring the ecosystem of web-based device
  fingerprinting.
\newblock In {\em IEEE Symposium on S\&P}, pages 541--555, May 2013.

\bibitem{Olejnik13}
L.~Olejnik, C.~Castelluccia, and A.~Janc.
\newblock On the uniqueness of web browsing history patterns.
\newblock {\em Annals of Telecommunications}, 69(1), February 2014.

\bibitem{Oliner:2013}
A.~J. Oliner, A.~P. Iyer, I.~Stoica, E.~Lagerspetz, and S.~Tarkoma.
\newblock Carat: Collaborative energy diagnosis for mobile devices.
\newblock In {\em ACM SenSys}, 2013.

\bibitem{Truong:2014}
H.~T.~T. Truong, E.~Lagerspetz, P.~Nurmi, A.~J. Oliner, S.~Tarkoma, N.~Asokan,
  and S.~Bhattacharya.
\newblock The company you keep: Mobile malware infection rates and inexpensive
  risk indicators.
\newblock In {\em WWW}, 2014.

\bibitem{Zang2011}
H.~Zang and J.~Bolot.
\newblock Anonymization of location data does not work: A large-scale
  measurement study.
\newblock In {\em MobiCom}, 2011.

\end{thebibliography}

%
%

\end{document}